%% file: main.tex
\documentclass[table]{article}
\usepackage[a4paper, margin=1.2in]{geometry}%
\usepackage{amsmath}
\usepackage{multirow}
\usepackage{authblk}
\usepackage{xcolor}
\usepackage{listings}
\usepackage{todonotes}
\usepackage[english]{babel}
\usepackage[inline]{enumitem}
\usepackage{graphicx}
\usepackage{csquotes}
\usepackage{tabularx}
\usepackage{booktabs}
\usepackage[T1]{fontenc}
\usepackage{xltabular}
\usepackage{longtable}
\usepackage{array}
\usepackage{subcaption}
\usepackage{pgf-pie}
\usepackage{amssymb}
\usepackage{xurl}
\usepackage{hyperref}

\usepackage{tikz}
\usetikzlibrary{arrows.meta,positioning,fit,calc,shadows,shapes.symbols,decorations.pathmorphing,backgrounds}
\definecolor{gdvBackground}{HTML}{f8f9fa}
\definecolor{gdvText}{HTML}{000000}
\definecolor{gdvKeyword}{HTML}{AF58BA}
\definecolor{gdvString}{HTML}{FF1F5B}
\definecolor{gdvComment}{HTML}{A0B1BA}     
\definecolor{gdvNumber}{HTML}{000000}
\definecolor{gdvFunction}{RGB}{111,66,193}

\definecolor{geoRed}{HTML}{ff1f5b}
\definecolor{geoGreen}{HTML}{00CD6C}
\definecolor{geoYellow}{HTML}{ffc61e}
\definecolor{geoOrange}{HTML}{f28522}
\definecolor{geoDBlue}{HTML}{0D4A70}
\definecolor{geoD2Blue}{HTML}{22639c}
\definecolor{geoMBlue}{HTML}{6cb0d6}
\definecolor{geoLBlue}{HTML}{E4f1f7}

\def\checkmark{\tikz\fill[scale=0.4](0,.35) -- (.25,0) -- (1,.7) -- (.25,.15) -- cycle;} 

\lstdefinelanguage{json}{
    string=[s]{"}{"},
    stringstyle=\color{gdvString},
    comment=[l]{:},
    commentstyle=\color{gdvComment},
}
\lstdefinelanguage{Markdown}{
  sensitive=true,
  morestring=[b]`,
  morecomment=[s]{<!--}{-->},
  alsoletter={\#},
  morekeywords={
    \#, \#\#, \#\#\#, \#\#\#\#, \#\#\#\#\#, \#\#\#\#\#\#,
    **, __, *, _, ~~,
    >, -, +,
    [, ], (, ), !,
  },
  morecomment=[s]{```}{```},
  alsoletter={:/},
}
\lstdefinestyle{geodataviz}{
    backgroundcolor=\color{gdvBackground},
    basicstyle=\ttfamily\scriptsize\color{gdvText},
    keywordstyle=\bfseries\color{gdvKeyword},
    stringstyle=\color{gdvString},
    commentstyle=\itshape\color{gdvComment},
    numberstyle=\tiny\color{gdvComment},
    breakatwhitespace=false,
    columns=flexible,
    numbers=left,
    numbersep=8pt,
    frame=single,
    rulecolor=\color{gdvComment},
    breaklines=true,
    breakatwhitespace=true,
    showstringspaces=false,
    tabsize=2,
    captionpos=b,
    keepspaces=true
}

\providecommand{\keywords}[1]{\textbf{\textit{Keywords:}} #1}

\definecolor{primaryBlue}{RGB}{41,74,122}
\definecolor{accentBlue}{RGB}{70,130,180}
\definecolor{neutralGray}{RGB}{95,100,110}
\definecolor{lightGray}{RGB}{200,205,210}
\definecolor{darkGray}{RGB}{60,65,75}
\begin{document}

\author[1]{George Tsigkourakos}
\author[1,2]{Constantinos Patsakis}
\affil[1]{Department of Informatics, University of Piraeus, 80 Karaoli \& Dimitriou str., 18534 Piraeus, Greece}
\affil[2]{Information Management Systems Institute, Athena Research Centre, Artemidos 6, Marousi 15125, Greece}
\date{}
\title{QRS: A Rule-Synthesizing Neuro-Symbolic Triad for Autonomous Vulnerability Discovery}
\maketitle

\begin{abstract}
Static Application Security Testing (SAST) tools are integral to modern DevSecOps pipelines, yet tools like CodeQL, Semgrep, and SonarQube remain fundamentally constrained, as they require expert-crafted queries, generate excessive false positives, and detect only predefined vulnerability patterns. Recent work has explored augmenting SAST with Large Language Models (LLMs), but these approaches typically require model fine-tuning or rely on LLMs to triage existing tool outputs rather than reason directly about vulnerability semantics.

We introduce QRS (Query, Review, Sanitize), a neuro-symbolic framework that inverts this paradigm, relocating the LLM to the pipeline's generative core. Rather than filtering results from static rules, QRS employs three specialized autonomous agents that generate CodeQL queries from a structured schema definition and few-shot examples, then validate findings through semantic reasoning and by synthesizing minimal proof-of-concept exploit candidates, which are verified manually before disclosure. This architecture allows QRS to contextualize vulnerability classes beyond predefined patterns while substantially reducing the volume of code that must be audited. We evaluate QRS on complete packages rather than isolated snippets, across both Python and Java ecosystems and three datasets. On the 100 most-downloaded PyPI packages, QRS reaches up to 94.06\% verdict accuracy and surfaces 41 medium-to-high vulnerabilities: 8 received new CVE identifiers, 4 were accepted as documentation acknowledgments, and the remaining 29 were previously published CVEs that QRS rediscovered from source alone. To evaluate language portability and enable direct comparison with prior work, we extend QRS to support Java and evaluate it on CWE-Bench-Java, detecting 58 of 90 in-scope CVEs (64.44\%) at 87.60\% accuracy, 98.79\% recall, 0.857 F1 (0.794 macro-averaged) score, in a single configuration scan.
QRS achieves these results with low time overhead and manageable token costs, while handling large codebases, demonstrating that LLM-driven query synthesis and code review can complement manually curated rule sets and surface vulnerability patterns that evade existing industry tools.
\end{abstract}


\section{Introduction}
Static Application Security Testing (SAST) tools have become a cornerstone of modern software security practices, embedded in secure Software Development Lifecycles (SDLCs) and DevSecOps pipelines across the industry. Tools such as CodeQL~\cite{CodeQL}, Semgrep/Opengrep~\cite{bennett2024semgrep,opengrep}, Bandit~\cite{bandit}, and SonarQube~\cite{SonarCloud} enable developers to identify potential vulnerabilities before code reaches production. However, these tools suffer from fundamental limitations that undermine their effectiveness in practice. For instance, traditional SAST operates by pattern-matching predefined rule sets, blinding it from the underlying application logic and semantic context. This creates a constant race between emerging vulnerability patterns and the rules designed to detect them. When rules are too narrow, complex vulnerabilities slip through undetected. On the contrary, when rules are too broad, developers face an overwhelming number of false positives that undermine trust in the tooling and consume valuable engineering time to triage. Thus, fatigue from the numerous false positives may lead developers to ignore the reported issues~\cite{8987507}. Customizing these tools to balance precision and recall requires significant expertise in query languages and deep familiarity with the target codebase, expertise that many development teams lack.

The security of the open-source software supply chain compounds these challenges. Modern applications depend heavily on third-party libraries, yet developers routinely trust popular packages without thorough review. This implicit trust is increasingly exploited: supply-chain attacks have surged in recent years, with malicious or vulnerable code propagating through widely used dependencies. The recent cases of XZ Utils~\cite{xz} and the Shai-Hulud worm~\cite{Shai_Hulud} are examples of open-source software compromise that led to the infection of thousands of hosts and repositories, despite the existence of SAST tooling. Thus, while SAST tools can scan dependencies, their rigid rule-based approach struggles to identify novel vulnerability patterns or context-dependent security issues in unfamiliar codebases.

Large Language Models (LLMs) offer a potential remedy through their capacity for semantic understanding and contextual reasoning. Recent work has explored the application of LLMs to vulnerability detection, taking advantage of their ability to comprehend code logic beyond surface-level patterns~\cite{lekssays2025llmxcpg,sun2024gptscan,guo2024outside,DBLP:conf/uss/FangMS0ZFATNWH24}. However, LLMs are fundamentally prediction-based systems, sophisticated pattern matchers that lack mechanisms for self-verification. Without being grounded in external tools or structured workflows, LLMs may hallucinate vulnerabilities, miss critical context, and produce inconsistent results. Directly prompting an LLM to review code yields unreliable findings that still require extensive human validation. Moreover, they heavily rely on training datasets~\cite{guo2024outside} and tend to degrade when evaluated on large codebases, burning tokens unmanageably.

In this work, we introduce QRS (\textbf{Q}uery, \textbf{R}eview, \textbf{S}anitize), a hybrid framework that combines the precision of SAST with the semantic reasoning capabilities of LLMs through a pipeline of three autonomous agents. Rather than replacing traditional tools or relying solely on LLM inference, QRS orchestrates both in an iterative, mutually reinforcing workflow pipeline. Unlike approaches that depend on task-specific model fine-tuning~\cite{lekssays2025llmxcpg} or on pre-existing vulnerability metadata~\cite{qlcoder2025}, QRS demonstrates that general-purpose LLMs can synthesize complex CodeQL logic using a lightweight schema definition and few-shot examples. The framework generates vulnerability-detection queries, eliminating the need for manually maintained rule sets, while preserving the rigor of static analysis and strengthening the final output through evidence-based reasoning. In this regard, the QRS framework decouples the vulnerability detection process into three specialized agents, each augmented with custom toolsets:

\begin{itemize}
    \item The \textbf{Query (Q) agent} synthesizes CodeQL queries from self-imposed hypotheses, after being injected with a knowledge document, translating high-level vulnerability descriptions into precise static analysis rules without requiring manual query authorship.
    \item The \textbf{Review (R) agent} performs a semantic reachability verification, analyzing the code context surrounding each potential finding, tracing data flows from source to sink, thoroughly assessing exploitability, and generating exploitation suggestions before submitting validated findings.
    \item The \textbf{Sanitize (S) agent} performs a final evidence-only assessment in a separate context window, eliminating false positives and consolidating results into high-confidence vulnerability reports, while assigning labels to each finding.
\end{itemize}
This architecture effectively mimics the workflow of an expert code auditor while maintaining low context windows, a bounded runtime, and manageable token costs, which are critical for practical deployment at scale.

To assess the efficacy of QRS, we evaluate our framework in three complementary settings. First, we evaluated the framework's functionality on 20 historical vulnerabilities in popular PyPI libraries using precompiled high-level strategy instructions, demonstrating that QRS can identify known vulnerability patterns. Second, we apply QRS to the 100 most popular PyPI packages as of November 2025, simulating real-world discovery conditions in heavily monitored source code. Third, we compile generic CWE definitions for 022, 078, 079, and 094 and scan the CWE-Bench-Java~\cite{cwe-bench-java} dataset, addressing a key limitation of prior evaluations that rely on incomparable datasets.

\noindent\textbf{Main contributions.}
We introduce QRS, a pipelined, configurable, and extensible multi-agent framework that demonstrates practical security applications beyond benchmark performance alone. Our evaluation of the 100 most popular PyPI packages identified 41 vulnerabilities, 8 of which we responsibly disclosed and received CVEs with CVSS scores ranging from 4.5 to 8.6, and 4 additional issues that were acknowledged by the maintainer updates on documentation. The newly discovered CVEs affected packages that reach 3 billion monthly downloads, are maintained by more than 1000 contributors, and are dependencies of more than 2 million repositories. The vulnerabilities persisted despite being scanned by GitHub's automated toolchain of CodeQL, Dependabot, and AI-enabled Copilot Autofix, and were patched swiftly, in most cases, within a few days. The findings span across 115 CWEs, including complex vulnerability classes, such as TOCTOU Race Conditions, ASN.1 octet memory exhaustion, Path Traversal chains, and Command Execution, demonstrating the framework's capability to detect exotic patterns that evade traditional static code analysis and security expert reviews. Contrary to the current state of the art, QRS uses LLMs to craft CodeQL queries and, using agentic loops, prunes noise, labels the results, prioritizes them, and explores exploitability while slicing the auditable code. Thus, we demonstrate that LLM agentic architectures can help discover complex real-world security issues at scale, including cases that existing automated tools miss. Note that all findings were manually reviewed and submitted accordingly.

\section{The QRS Methodology}
\label{sec:framework}
\subsection{Scope}
QRS targets a practical threat scenario end-to-end, identifying exploitable vulnerabilities in source code before attackers discover and weaponize them in production. Thus, we assume defenders have access to the source code and seek to minimize manual review effort while maximizing detection of genuine security issues. Moreover, we assume that defenders have access to LLMs they use to craft CodeQL queries and to conduct code reviews. Note that all CodeQL queries are executed locally, and LLMs parse code slices, rather than the entire codebase. Therefore, the codebase is substantially reduced into objects and not fully exposed to an LLM. Moreover, we do not evaluate adversarial code intentionally crafted to evade LLM-guided query synthesis, which remains an important direction for future work. As such, our experimental setup does not address: \begin{enumerate*}[label=(\roman*)]
\item vulnerabilities in compiled extensions or native code, \item logic vulnerabilities requiring dynamic analysis, \item supply-chain attacks at the package distribution level (e.g., compromised PyPI uploads), and \item adversarial evasion where attackers specifically craft code to bypass QRS detection.\end{enumerate*} Since there are tools to efficiently determine whether a specific library is known to be vulnerable, we do not consider imported libraries whose code is not provided in the package. Note that QRS is restricted to CodeQL-supported languages~\cite{codeql_languages}.

These exclusions reflect practical scoping decisions rather than fundamental limitations and are widely adopted by similar software security products. Compiled extensions and native code require binary analysis techniques outside CodeQL's source-level approach. Logic vulnerabilities that necessitate dynamic analysis constitute a complementary threat class best addressed by fuzzing or runtime testing. Distribution-level supply-chain attacks (e.g., compromised PyPI uploads) occur outside the source code and require monitoring of package repositories rather than code analysis. Finally, while adversarial evasion remains theoretically possible, our evaluation against heavily scrutinized production packages suggests that current obfuscation techniques in legitimate codebases do not significantly impede detection. These boundaries align with standard SAST threat models and do not diminish QRS's utility for its intended use case: identifying exploitable vulnerabilities in source code prior to deployment.

\subsection{Overview}

QRS is a configurable, self-healing, rule-synthesizing triad of autonomous neuro-symbolic agents that collaborate in a pipelined manner to detect and validate security vulnerabilities. The agents are separate in terms of context and custom tools, and each subsequent agent requires an artifact generated by the previous one, except the entrypoint \textbf{Q} agent. The framework is model-agnostic; it unifies model APIs via LiteLLM~\cite{litellm}, providing compatibility with Gemini, Claude, GPT, and DeepSeek variants. The architecture balances contextual understanding with computational efficiency through a hybrid approach combining custom-built tooling with LLM-based reasoning and AI agents. 

QRS can handle any supported CodeQL language by swapping the target language's QL knowledge predicates, without requiring architectural re-engineering. Our experimental setup in \autoref{sec:validation} focuses on complete Python and Java packages, and a set of distinct configurations; nonetheless, it can be easily ported to support other programming languages. To extend functionality, an update of knowledge is sufficient. Additionally, QRS can be configured to run with a set of attributes per agent, including model, temperature, iterations, maximum output tokens, and other generic caching attributes, as needed.

\begin{figure*}[!ht]
    \centering
    \resizebox{.9\textwidth}{!}{
    \input{architecture}
    }
    \caption{Overview of the QRS framework architecture.} 
    \label{fig:qrs}
\end{figure*}

\autoref{fig:qrs} illustrates the framework's architecture. The three core components are: the \textbf{Q agent} (hypothesis and query generation and refinement engine), which self-synthesizes CodeQL queries from documentation; the \textbf{R agent} (review, exploitation suggestion and reasoning engine), which performs semantic verification of potential findings by analyzing sources, sinks and code flows; and the \textbf{S agent} (arbitration and finding sanitization engine), which renders final verdicts, filters false positives and assigns multiple labels to findings. Each agent operates with a dedicated toolset and iterates up to a user-defined limit. Custom static pre-processing utilities execute before each scan to clone, prepare paths, and construct the relevant CodeQL database. All findings undergo manual expert review for final acceptance or rejection.

Each scan can be configured to involve any combination of models, iterations, maximum tokens, and temperature variations. For brevity, the model combinations remained within the same family. A `big' model was always used to craft queries and the same or `smaller' ones to review code or produce proof-of-concept (PoC) exploitation snippets. The same configuration for maximum iterations, output tokens, and temperatures of 0 and 1 has been tested for each dataset. We evaluate temperature settings of 0 and 1 because temperature behavior varies across model families, and these values represent reasonable bounds for code generation and review tasks with negligible hallucinations.

All agents import a structured markdown input and an optional instruction set for CWE specifics in their \texttt{system\_prompt} with a declarative Role (\emph{who the agent is}), Goal (\emph{what is its purpose}), Tools (\emph{what are its new skills}), Strategy (\emph{how to tackle tasks}), and \emph{Examples} when necessary. From this point on, we will refer to
\begin{enumerate*}[label=(\roman*)]
    \item \textbf{Main definition}: the Role, Goal, Tools, and Examples,
    \item \textbf{Knowledge definition}: the CodeQL query syntax knowledge base,
    \item \textbf{Extra strategy definition}: the CWE or focus points preprompts.
\end{enumerate*}

\subsection {Query Generation}
\label{subsec:q-agent}

The \textbf{Q} agent serves as the framework's primary guiding and slicing engine. \textbf{Q} gets injected with the \textbf{main} and \textbf{knowledge} and optionally with the \textbf{extra strategy} definitions. The agent is tasked with translating self-generated security hypotheses into executable CodeQL logic. Unlike traditional SAST, \textbf{Q} operates as a feedback-loop query synthesis engine that dynamically constructs and refines queries tailored to the library's context and previous outputs, while assigning severity levels and CWEs to each hypothesis, up to user-specified iterations. The main workflow is governed by the following modules: \emph{Retrieval-Augmented QL Schema}, \emph{Structural Library Reconnaissance}, and \emph{Feedback-Loop Query Refinement} with self-healing capabilities.

\noindent\textbf{Retrieval-Augmented QL Schema.}
The absence of predefined query rules requires a compact \texttt{knowledge.md} file, which serves as a sparse schema definition. A separate file for each language (Python and Java) contains CodeQL syntax primitives and canonical patterns, as seen in Appendix \autoref{lst:codeql-knowledge}.

\noindent\textbf{Structural Library Reconnaissance.}
To address the `cold-start' problem, inherent in zero rule knowledge analysis, the \textbf{Q} agent is instructed to run a custom package profiling utility \textbf{once}, consuming the first iteration. This lightweight static scan utility captures file paths, line numbers, import dependencies, and package metadata. This generates a JSON metadata artifact of the scanned library and feeds it back to the agent, as shown in Appendix \autoref{lst:package_profile}.

\noindent\textbf{Feedback-Loop Query Refinement.}
After completing the first iteration, the agent has enough data to begin self-generating hypotheses and CodeQL queries. Each executed query's output is fed back into the context, allowing the agent to self-refine or continue on another route via the \texttt{execute\_query} tool. CodeQL's declarative syntax is strict, often leading to compilation failures. These failures are also fed back to the context, allowing the agent to decide whether to enter a recovery \texttt{fixup\_query} subroutine (see \autoref{lst:q-agent-iteration} in Appendix) for:
\begin{enumerate*}[label=(\roman*)]
    \item \textbf{Isolation}: The failed raw query and compilation errors (stderr) are moved to a dedicated, ephemeral isolated thread, preventing `context pollution' and ensuring focus on the same hypothesis, severity, and CWE targets.
    \item \textbf{Refinement}: The agent attempts to patch the failed query's logic based on each error in an iterative manner.
    \item \textbf{Constraint}: This separate agentic loop is bounded by a retry budget (default: 3 attempts). If the query still fails after budget exhaustion, the results are returned to the main thread, forcing the agent to either change approach or entirely discard this hypothesis.
\end{enumerate*}

Generated queries, as shown in \autoref{lst:dangerous-exec} in the Appendix, are executed locally against the CodeQL database and are sent back to the \textbf{Q} agent. Successful queries and respective findings are then formatted as a navigational map for the downstream Review (\textbf{R}) agent using the \texttt{finalize\_report} tool in JSON/SARIF format.

\subsubsection{Extra Strategy Directives}

To accommodate targeted audit requirements or specific bug-hunting scenarios, the \textbf{Q} agent implements extra natural language strategy input directives. This mechanism allows researchers to inject additional high-level directives via \texttt{preprompt.md} files, see \autoref{lst:q-priority-areas} in the Appendix, and instruct the agent to focus on specific types of vulnerabilities, or even set restrictions and adjust the weights of each scan. This way, the framework can be extended into a broader-spectrum scanner that bridges the gap between human intuition and automated analysis. Both \textbf{Q} and \textbf{R} agents may include these extra directives.
For the Historical 20 and CWE-Bench-Java evaluations, we develop strategies that cover specific focus points or CWE templates, respectively, for each set. For the Top 100, we skip this input so the agents can freely decide what to look for, given the package metadata and query outputs, as we will see later.

\subsection {Semantic Review \& Exploitation}
\label{subsec:r-agent}

While the \textbf{Q} agent can identify vulnerability patterns based on structural logic, it lacks the semantic depth and raw code inputs needed to determine exploitability. The \textbf{R} agent is a semantic verification and evidence-gathering engine that acts as an investigator, looking at whether a code slice can actually be exploited or whether sanitizers, mitigators, or externally unreachable paths exist. Since \textbf{Q} provides a report of potentially interesting findings, \textbf{R} consumes this report in a separate context window and tries to identify vulnerable points in the codebase end-to-end. \textbf{R} gets also injected with the \textbf{main} and optional \textbf{extra strategy} definitions to complete the task at hand. No \textbf{knowledge} is needed, as the workflow here is based on four components: \emph{MITRE Taxonomy-Driven Clustering}, \emph{Feedback-Loop Context Reconstruction}, \emph{Evidence \& Reasoning judgment}, and \emph{Exploitation Suggestion}. The output of these tasks is handed over in JSON/SARIF format to the downstream Sanitization (\textbf{S}) agent after all verdict submissions have been completed.

\noindent\textbf{MITRE Taxonomy-Driven Clustering.}
Raw query results without semantics can often lead to context window degradation or the omission of more complex scenarios. As coding patterns and anti-patterns tend to recur, findings are statically clustered with \texttt{group\_findings\_by\_mitre}, before \textbf{R}'s init, by relevance according to MITRE's CWE database hierarchy~\cite{cwe_database}. By pivoting from file locations into groups of findings, we change \textbf{R}'s viewpoint to a taxonomy-driven review, based on severity, prevalence, likelihood, CWE ID, and related CWE IDs. Each group is also minified to remove duplicates in file locations, code flows, and unnecessary fields. Each cluster is then processed in an isolated context thread, with its own user-defined iteration budget set to $I \times N_{findings}$, ensuring thread context separation per group. This may lead to duplicate vulnerability entries, since the same finding may appear multiple times in \textbf{Q}'s queries and be mutated in various cluster definitions. The final duplicates are statically eliminated before being handed over to \textbf{S}, based on the CWE, file, and location of the pinpointed vulnerability.

\noindent\textbf{Feedback-Loop Context Reconstruction.} The starting point of \textbf{R} agent. Unlike a traditional SAST tool that alerts a developer with static snapshots of findings, the \textbf{R} agent interrogates the codebase to reason and validate. Two semantic-slicing tools are used for this purpose. First, \texttt{trace\_flow} maps the flow of a propagation path from the source to the sink with a line tolerance limit of 1 to 3 surrounding lines. Beneath this tool is a single parameterized, hard-coded CodeQL predicate query with sink-to-source and tolerance parameters. Then, \texttt{grep\_search} is a sandboxed local grep search capability. It enables the agent to perform a pattern-enabled grep under a restricted local path.
Bounded raw code snippet extraction via \texttt{extract\_code\_snippet} seems to be the most preferred tool of the tested AI models, as we will see later in \autoref{sec:qrs-metrics} and \autoref{sec:ablation}. 

\noindent\textbf{Evidence \& Reasoning judgment.}
To enhance the traditional binary True Positive / False Positive verdict system, we introduce a more nuanced evidence-based classification. The \textbf{R} agent is required to provide traditional SAST fields, CWE, alongside reasoning, code context, and data flow analysis, forcing the agent to prove the verdict claim as a:
\begin{enumerate*}[label=(\roman*)]
    \item \emph{True Positive (TP) (Confidence $\geq 90\%$)}: A vulnerable code pattern finding, with a high degree of confidence. Requires generating a minimal exploit snippet, a scenario, the prerequisites to enable the attack, the likelihood, the impact, and the exploitability level.
    \item \emph{Manual Review (MR) (Confidence $\geq 50\%$)}: A TP, but the complexity of the exploit chain or the agent's decision horizon deems it unable to reach certainty.
    \item \emph{False Positive (FP)}: A non-vulnerable code pattern.
\end{enumerate*}
The MR label is a guide for a code auditor, as the confidence binary split is evenly distributed between 0--49 and 50--100 and, as such, will be treated as a TP.

By forcing the model to generate the above, we leverage the chain-of-thought effect~\cite{claudereasoning, geminireasoning, gptreasoning, deepseekreasoning}, allowing the agent to articulate a plausible exploitation path.

\noindent\textbf{Exploitation Suggestion.}
To bridge the gap between theoretical reachability and practical exploitability, the \textbf{R} agent must strongly verify practical exploitability. Once a vulnerability candidate is semantically verified and ready to be submitted as a finding, the agent plays an additional adversarial role and must synthesize a minimal working PoC exploit. This involves constraint solving; the prerequisites to materialize the attack scenario, artifact construction; the payload, and a semantic verification, a scenario for the attack, including the scenario's category (e.g., Remote code execution), likelihood (low-high), impact (low-high), and exploitability (low-high); see \autoref{lst:r-agent-finding} and \autoref{lst:yaml-exploit} in the Appendix. As this capability was sufficient to guide manual reviews and might also get blocked by frontier models, we limited the scope to simple PoC generation on this setup while enabling future research.

\subsubsection{Extra Strategy Directives}

Similarly to \textbf{Q}, the \textbf{R} agent implements a natural language input parameter, allowing the injection of an extra high-level directive via a \texttt{preprompt.md} strategy; see \autoref{lst:r-review-priorities} in the Appendix, and guides the agent to look for specific types of vulnerabilities, CWEs, TP or FP indicators, and critical focus areas, adjusting weights of each review process.

\subsection {Sanitization \& Verdict}

The final component of the neuro-symbolic triad is the minimal \textbf{S} agent. This agent is designed to function as an independent judge of evidence over previously generated artifacts rather than a detector, serving as a correction and hallucination-avoidance mechanism. \textbf{S} receives \textbf{R}'s complete artifact and iterates over each reported vulnerability in a separate context window, adding a final verdict on what the other two `reported'. Compared to the other two agents that run their tools according to their preferences, the \textbf{S} agent populates context in an instructed series and iterates over all \textbf{R} findings.

Agent \textbf{S} is modular, minimal, and detachable from the workflow. It reduces noise as an evidence-bounded review protocol and assigns taxonomies to each finding, strengthens or weakens \textbf{R}'s decisions, and adds insights for expert analysts through three verification layers: \emph{Novelty Verification}, \emph{Operational Analysis}, and \emph{Vulnerability Classification}.

\noindent\textbf{Novelty Verification.}
To distinguish between known technical debt and genuine zero-day threats, a \texttt{search\_cve} retrieval tool that directly queries the Common Vulnerabilities and Exposures (CVE) database~\cite{cve_database} is invoked statically before the agent starts. The output is then submitted to the agent to determine whether the issue is a repetition of a published CVE or a novel finding. To avoid direct CVE correlation with the examined finding, the tool is called after the review phase has completed, and the agent simply assigns a label.

\noindent\textbf{Operational Analysis.}
Frequent sources of FPs in static analysis are: the absence of context, API reachability (e.g., the vulnerability is in deprecated or internal-only code), and non-production code (e.g., the vulnerability is in a test suite for the actual library). The agent analyzes the submitted surrounding logic, code flows, and file paths, respectively, to determine the operational scope of each finding and updates each verdict via \texttt{verify\_finding}.

\noindent\textbf{Vulnerability Classification.}
Before finalizing, the agent applies a multi-label tagging taxonomy to move further away from the binary TP/FP result and adds actionable intelligence concerns. The available flags are
\begin{enumerate*}[label=(\roman*)]
    \item \texttt{KNOWN\_CVE}: already published CVE, matching the finding;
    \item \texttt{ZERO\_DAY}: a finding with no matching CVE, high to critical severity, and relatively easy to exploit;
    \item \texttt{NOVEL\_FINDING}: a finding with no matching CVE, medium to high severity, and exploitability uncertainty;
    \item \texttt{BUILD\_SCRIPT}: a finding found in a build script;
    \item \texttt{TEST\_CODE}: a finding found in test/example/smoke code;
    \item \texttt{EXPLOITABILITY}: a vulnerability with exploitability and prerequisite concerns in context;
    \item \texttt{HALLUCINATION}: the claimed code context does not match the finding;
    \item \texttt{CODE\_SMELL}: not an actual vulnerability but a best-practice issue, and
    \item \texttt{CONTEXT\_MISMATCH}: the finding does not align with the code context.
\end{enumerate*}
Upon completion, each finalized verdict, as seen in \autoref{lst:s-artifact-output}, is appended to \textbf{R}'s respective artifact, as seen in \autoref{lst:r-agent-finding}, presenting an enhanced and curated JSON/SARIF output to the expert security analyst.

\section{Framework validation}
\label{sec:validation}

According to the TIOBE index~\cite{tiobe}, Python is the most popular language. The Python Package Index (PyPI), a centralized repository of open-source Python packages, is our data source. Using PyPI as a reference point, we can establish an experimental dataset and a real-world setting to maximize the potential impact of this research. To demonstrate the effectiveness of QRS in other programming languages, we also evaluate it on Java, which ranks fourth on the TIOBE index.

We initially evaluate QRS on three datasets, two focused on Python and one on Java. First, we use a manually curated Historical 20 (Hist20) dataset to establish the framework's functionality and detection capability; second, a real-world production dataset on the Top 100 (Top100) PyPI packages; and finally, on a trimmed version of CVE-Bench-Java, a benchmark dataset
of 120 previously vetted CVEs in real-world Java projects. In \autoref{sec:observations}, we compare QRS with known SAST and run a single benchmark scan configuration with CWE-Bench-Java.

\noindent\textbf{Framework configuration.}
Since QRS can run with any model combination, we prefer higher-reasoning models for the \textbf{Q} agent and lower for \textbf{R} and \textbf{S} to limit token consumption on less intensive tasks for the Python datasets. For Java packages, we use the same configuration and model across the pipeline. To limit the scope of the experiments, we keep each configuration within the same model family and assign temperature values of 0 and 1, corresponding to deterministic and creative sampling, respectively. We annotate models in short with their respective temperature configuration subscripts as explained in the Appendix \autoref{tab:model-configuration-matrix}.

\subsection{Dataset Construction}

To collect the two Python datasets, we used the official Python Index Project~\cite{pypi} and a snapshot of the most downloaded PyPI packages from the Open-Source project `Top PyPI Packages'~\cite{toppypipackages}, as of November 2025. We implemented an automated crawler to retrieve archived source distributions from PyPI's repository for direct analysis. For the prior work benchmark, we use the available dataset of package names~\cite{cwe-bench-java} and clone each vulnerable version locally. Note that not all Java packages were available, and that several vulnerabilities were not in Java code, so CWE-Bench-Java was trimmed down to 90 CVEs. We note that CWE labels were taken from the upstream GitHub Advisory entry rather than from IRIS's~\cite{iris2025} manual re-labeling. Our final in-scope Java dataset is restricted to the same four CWE classes IRIS evaluates. No further processing was performed, since QRS accepts archived packages and handles extraction and CodeQL database generation.

\subsection{Metrics Definitions}

Vulnerability detection in our datasets lacks the binary labels of curated lab benchmarks. For Top100 datasets, ground truth is established through expert review and maintainer feedback. For Hist20 and CWE-Bench-Java, the ground truth is externally provided, as each CVE includes fixed files, classes, and methods that are manually vetted by its authors. We define metrics as
\begin{enumerate*}[label=(\roman*)]
    \item \textbf{True Positive (TP)/Manual Review (MR)}: The finding describes a genuine or almost certain vulnerability. Findings with too many prerequisites are considered TP, since the code snippet is vulnerable, but includes exploitability concerns, in agreement with expert review.
    \item \textbf{False Positive (FP)}: The finding describes a non-vulnerable code snippet. Either the attack vector is infeasible or mitigated by the runtime context.
    \item \textbf{True Negative (TN)}: The finding describes a non-vulnerable code snippet, in agreement with expert review, and
    \item \textbf{False Negative (FN)}: The finding describes a missed vulnerability.
\end{enumerate*} To calculate metrics, we use the confusion matrix of \autoref{tab:confusion-matrix}. Note that our metric definition makes them objective and rigorous. For example, we do not simply declare a new vulnerability and mark it as positive; it is verified by the package owners and published as a CVE. Thus, although the ground-truth labeling is more limited, especially for Top100, the results remain subject to expert judgment and maintainer feedback, introducing unavoidable subjectivity despite our efforts to standardize validation. However, this approach does not inflate our results.

\begin{table}[ht]
\centering
\caption{Verdict-retention confusion matrix: S Agent vs Human (H) Verdict}
\label{tab:confusion-matrix}
\scriptsize
\rowcolors{2}{gray!10}{white}
\begin{tabular}{@{}lcl@{}}
\toprule
\textbf{Classification} & \textbf{S agent $\rightarrow$ Human} & \textbf{Interpretation} \\
\midrule
TP & TP/MR $\rightarrow$ TP & S + H correct acceptance \\
FP & TP/MR $\rightarrow$ FP & S incorrectly flagged \\
TN & FP $\rightarrow$ FP & S + H correct rejection \\
FN & FP $\rightarrow$ TP & S missed vulnerability \\
\bottomrule
\end{tabular}
\end{table}

\subsection{Historical 20 Dataset}
\label{subsec:hist20}

To establish functionality, we curate a dataset of 20 PyPI package versions with reported CVEs, containing 20 CVEs spanning 2018--2025 across 12 CWE categories (see \autoref{tab:hist20-composition}). Each package contains at least one CVE with a published advisory right before patch application. Due to the limited sample size (N=20) and potential leak of known CVEs into the training datasets of models, metrics on the Hist20 dataset should be interpreted with caution, as this is a framework verification dataset.

\begin{table}[!ht]
\centering
\caption{Hist20: Known CVE Composition}
\label{tab:hist20-composition}
\scriptsize
\begin{tabular}{@{}llcl@{}}
\toprule
\textbf{Package} & \textbf{CVE} & \textbf{CVSS} & \textbf{Vulnerability Type} \\
\midrule
\rowcolor{gray!10}
aiohttp-3.8.5 & CVE-2023-47627 & 7.5 & HTTP Req. Smuggle \\
\rowcolor{white}
aiohttp-3.9.1 & CVE-2024-23334 & 7.5 & Path Traversal \\
\rowcolor{gray!10}
cryptography-39.0 & CVE-2023-23931 & 4.8 & Memory Corruption \\
\rowcolor{white}
django-4.1 & CVE-2023-31047 & 9.8 & Validation Bypass \\
\rowcolor{gray!10}
django-4.2 & CVE-2023-36053 & 7.5 & ReDoS \\
\rowcolor{white}
fastapi-0.65.1 & CVE-2021-32677 & 8.1 & CSRF \\
\rowcolor{gray!10}
jinja2-3.1.2 & CVE-2024-22195 & 6.1 & XSS \\
\rowcolor{white}
jinja2-3.1.5 & CVE-2025-27516 & 5.4 & Sandbox Escape \\
\rowcolor{gray!10}
lxml-4.2.4 & CVE-2018-19787 & 6.1 & XSS \\
\rowcolor{white}
lxml-4.6.2 & CVE-2021-28957 & 6.1 & XSS \\
\rowcolor{gray!10}
pillow-8.0 & CVE-2022-22817 & 9.8 & Arbitrary Code Exec. \\
\rowcolor{white}
pillow-10.1 & CVE-2023-50447 & 9.8 & Arbitrary Code Exec. \\
\rowcolor{gray!10}
pydantic-1.8 & CVE-2021-29510 & 7.5 & DoS \\
\rowcolor{white}
pydantic-2.3 & CVE-2024-3772 & 5.9 & ReDoS \\
\rowcolor{gray!10}
pyyaml-5.3 & CVE-2020-14343 / CVE-2020-1747 & 9.8 & Arbitrary Code Exec.\\
\rowcolor{white}
requests-2.19 & CVE-2018-18074 & 7.5 & Info Disclosure \\
\rowcolor{gray!10}
requests-2.30 & CVE-2023-32681 & 6.1 & Info Disclosure \\
\rowcolor{white}
setuptools-69.1.1 & CVE-2024-6345 & 8.8 & Remote Code Exec. \\
\rowcolor{gray!10}
urllib3-1.26.17 & CVE-2023-45803 & 4.2 & Info Disclosure \\
\rowcolor{white}
werkzeug-2.2.2 & CVE-2023-23934 & 2.6 & Cookie Injection \\
\bottomrule
\end{tabular}
\end{table}

To assess functionality and efficacy, QRS ran with 4 basic configurations (C$_0$, G$_0$, GPT$_0$, DR$_0$), included natural language instructions, and focused only on a subset of CWEs, with the CVE search tool disabled to avoid direct correlation. QRS detected 19/20 CVEs across Hist20 packages in \autoref{tab:hist20-composition}, as seen in \autoref{tab:hist20-model-detection}.

\begin{table}[!ht]
\centering
\caption{Hist20: CVE Detection}
\label{tab:hist20-model-detection}
\scriptsize
\rowcolors{2}{gray!10}{white}
\begin{tabular}{@{}lccccc@{}}
\toprule
\textbf{CVE} & \textbf{C$_0$} & \textbf{DR$_0$} & \textbf{G$_0$} & \textbf{GPT$_0$} & \textbf{Consensus} \\
\midrule
CVE-2023-47627 & \checkmark & \checkmark & \checkmark & \checkmark & 4/4 \\
CVE-2024-23334 & \checkmark & & \checkmark & & 2/4 \\
CVE-2023-23931 & & \checkmark & \checkmark & & 2/4 \\
CVE-2023-31047 & \checkmark & & & & 1/4 \\
CVE-2023-36053 & & & \checkmark & & 1/4 \\
CVE-2021-32677 & & & & & 0/4 \\
CVE-2024-22195 & & \checkmark & \checkmark & \checkmark & 3/4 \\
CVE-2025-27516 & & \checkmark & & \checkmark & 2/4 \\
CVE-2018-19787 & \checkmark & \checkmark & \checkmark & \checkmark & 4/4 \\
CVE-2021-28957 & & & \checkmark & & 1/4 \\
CVE-2022-22817 & \checkmark & \checkmark & \checkmark & \checkmark & 4/4 \\
CVE-2023-50447 & \checkmark & \checkmark & \checkmark & \checkmark & 4/4 \\
CVE-2021-29510 & \checkmark & \checkmark & \checkmark & \checkmark & 4/4 \\
CVE-2024-3772 & \checkmark & \checkmark & \checkmark & & 3/4 \\
CVE-2020-14343 & \checkmark & \checkmark & \checkmark & \checkmark & 4/4 \\
CVE-2018-18074 & \checkmark & \checkmark & \checkmark & \checkmark & 4/4 \\
CVE-2023-32681 & & \checkmark & & & 1/4 \\
CVE-2024-6345 & \checkmark & \checkmark & \checkmark & \checkmark & 4/4 \\
CVE-2023-45803 & \checkmark & \checkmark & \checkmark & & 3/4 \\
CVE-2023-23934 & & \checkmark & \checkmark & & 2/4 \\
\midrule
\textbf{Total} & \textbf{12/20} & \textbf{15/20} & \textbf{16/20} & \textbf{10/20} & \textbf{2.7/4} \\
\bottomrule
\end{tabular}
\end{table}

\begin{table}[!ht]
\centering
\caption{Hist20: Prediction Performance Metrics}
\label{tab:hist-20-performance}
\scriptsize
\rowcolors{2}{gray!10}{white}
\begin{tabular}{@{}lccccc@{}}
\toprule
\textbf{Config} & \textbf{Correct/Predict} & \textbf{Accuracy} & \textbf{Precision} & \textbf{Recall} & \textbf{F1} \\
\midrule
C$_0$ & 102/118 & 86.44\% & 75.76\% & 75.76\% & 0.758 \\
DR$_0$ & 29/32 & 90.62\% & 86.96\% & 100.00\% & 0.930 \\
G$_0$ & 57/67 & 85.07\% & 69.70\% & 100.00\% & 0.821 \\
GPT$_0$ & 69/93 & 74.19\% & 44.12\% & 75.00\% & 0.556 \\
\bottomrule
\end{tabular}
\end{table}

\subsection{The Top 100 Dataset}
\label{subsec:top100}

After initial validation on Hist20, we evaluate the framework on real-world discovery capabilities. We collect the 100 most popular PyPI packages, but unlike Hist20, this dataset does not include a priori knowledge or \textbf{extra strategy} definitions, and there is no potential leak of unknown CVEs into the training data of each model. The package selection criteria are \emph{Popularity} (highest monthly download counts) and \emph{Maintenance}, with packages actively maintained to ensure responsible disclosure. Nevertheless, it may underestimate detection rates on less-maintained codebases where vulnerabilities are more likely to persist. As a result, the packages in our dataset are very popular, with monthly downloads ranging from 144 million (snowflake\_connector\_python-4.0.0) to 1.4 billion (boto3-1.40.64). Almost all packages (96\%) have a GitHub repository, with contributors ranging from 3 (typing\_inspection-0.4.2) to 892 (pygments-2.19.2), and commits ranging from 37 (mdurl-0.1.2) to 57.9k (grpcio-1.76.0), indicating that the packages are very active.

We evaluate QRS in an open-world dataset across 10 configurations (5 model families $\times$ 2 temperature settings $\times$ 100 packages). We allow the framework to operate autonomously, without predefined CWE and focus-point targets or specific CodeQL queries. At the time of writing, QRS identified 29 previously known vulnerabilities, 8 newly published CVEs, and received 4 acknowledgments with documentation updates.

\noindent\textbf{29 Previously known vulnerabilities.}
QRS successfully rediscovered the precise vulnerable code locations for 29 previously reported CVEs across this dataset. These vulnerabilities span packages that have either been patched in subsequent releases or remain documented as unsafe API usage patterns. Note that the majority of these vulnerabilities are moderate to critical; 21 have a CVSS score of at least 7.5. QRS also produced accurate verdicts distinguishing exploitable from mitigated code and labeled each finding accordingly (see \autoref{tab:top100-cve-detection}). It should be noted that the rediscovery was not due to any model training or to search engine queries from the LLM, as it was identified through each agent's tooling capabilities. \textbf{Q}'s CodeQL queries pointed out potential code snippets of interest, \textbf{R} traced the code flow through the review phase, and finally \textbf{S} utilized the \texttt{search\_cve} tool and assigned the corresponding \texttt{KNOWN\_CVE} label alongside the CVE ID.

\noindent\textbf{8 Novel vulnerabilities.}
QRS identified 8 previously unreported vulnerabilities across 7 packages. These vulnerabilities are complex in nature, e.g., Time-of-Check Time-of-Use (TOCTOU) Race Conditions, Denial-of-Service (DoS) Memory exhaustion, Path traversal via dependencies, Deserialization and Code Execution, and were missed by maintainers, security researchers, or automated tooling, including GitHub's CodeQL queries. These packages collectively reach 3 billion monthly downloads, are maintained by more than 1000 contributors, and are dependencies of more than 2 million other repositories. All vulnerabilities were responsibly disclosed to the maintainers and were patched promptly. The resulting CVEs are:

\noindent\textbf{CVE-2025-68146 | CWE-59,362,367 | Moderate - 6.3 | Race Condition in filelock < 3.20.0}. The package (351M monthly downloads, 55 contributors, 696k dependencies) provides widely-used platform-independent file locking mechanisms, used extensively in build systems, virtual environments, and a critical dependency for major projects such as PyTorch~\cite{pytorch}, Poetry~\cite{python_poetry}, and HuggingFace Transformers~\cite{hugghing_face_transformers}. QRS identified a local TOCTOU issue affecting locking mechanisms on both Linux and Windows platforms, which could corrupt or destroy files via symlink manipulation, exfiltrate sensitive metadata, or cause a Denial-of-Service (DoS). As a workaround, the maintainers suggested soft-locking mechanisms.

\noindent\textbf{CVE-2026-22701 | CWE-59,362,367 | Moderate - 5.3 | Race Condition in filelock < 3.20.3}. A follow-up disclosure after patching CVE-2025-68146 on soft-locking mechanisms. The local TOCTOU issue affected soft-locking mechanisms through symlink manipulation, leading to inconsistent states and DoS attacks.

\noindent\textbf{CVE-2026-22702 | CWE-59,362 | Moderate - 4.5 | Chained Race Condition in virtualenv < 20.36.1}. One of the most popular virtual environment isolation dependencies for Python (273M monthly downloads, 151 contributors, 493k dependencies). QRS highlighted a chained exploitation path that combines 2 findings in a 3-stage attack scenario via environment variables. In short, the flow included symlink manipulation in the application's data path, a lock-race condition, and corruption of data in locally shared temporary directories, leading to cache poisoning, information disclosure, and a lock-starvation DoS.

\noindent\textbf{CVE-2026-23949 | CWE-22 | High - 8.6 | Zip Slip Path Traversal in jaraco.context 5.2.0 - 6.1.0 affecting setuptools-80.9.0}. A cross-package dependency vulnerability, tracked in \texttt{setuptools} (880M monthly downloads, 546 contributors, 806k dependencies) using the vendored \texttt{jaraco.context}~\cite{jaraco_context, setuptools_vendored} (91M monthly downloads, 18 contributors, 14k dependencies) utility package, effectively affecting almost the entire PyPI ecosystem. QRS identified the flow of context across Python versions during archive extraction and traced the zip-slip Path Traversal vulnerability, introduced between versions, which allowed `../` sequences to bypass extraction filters and extract outside intended directories. This was affected by remote resources and could result in overwriting system files, information leakage, and a potential DoS.

\noindent\textbf{CVE-2026-23490 | CWE-770 | High - 7.5 | ASN.1 Memory Exhaustion in pyasn1 < 0.6.1}. A foundational package for ASN.1 parsing (367M monthly downloads, 13 contributors, 18k dependencies), used in cryptographic protocols and certificate validation. QRS highlighted that a malformed remote REL-OID (Relative Object Identifier) could trigger unbounded recursion during parsing, leading to memory exhaustion. Any service accepting a specially crafted ASN.1 payload, while using the package, is a potential victim of a severe remote DoS attack.

\noindent\textbf{CVE-2026-34993 | CWE-502 | High - 7.3 | Deserialization of untrusted Data in aiohttp via Cookiejar < 3.14}. A well-known async HTTP client/server framework (280M monthly downloads, 804 contributors, 752k dependencies). QRS detected a Deserialization scenario that enables local Code Execution via a shared cookiejar (e.g., shared container storage), via the \texttt{Cookiejar.load()} method that wraps around a \texttt{pickle.load()} implementation.

\noindent\textbf{CVE-2026-28684 | CWE-59,61 | Moderate - 6.6 | Symlink allows arbitrary file overwrite via cross-device fallback in python-dotenv < 1.2.1}. The standard Python utility that loads key-value pairs from a local \texttt{.env} file (273M downloads, 109 contributors, 1.6M dependencies). QRS traced a complex vulnerability that enables an attacker to pre-place a symlink and overwrite the victim's \texttt{.env} file when \texttt{set\_key|unset\_key} methods are called. This introduced DoS and integrity violations, while enabling privilege escalation and lateral movement when the caller runs in an elevated privilege context.

\noindent\textbf{CVE-2026-7246 | CWE-78 | High - 7.2 | Command Injection via shell metacharacters in click < 8.3.1}. A popular Python Command Line Interface creation toolkit (483M downloads, 421 contributors, 2.3M dependencies). QRS identified a double-quote shell-escape supply chain attack, enabling direct Command Execution in supply chain environments.

\noindent\textbf{4 Acknowledgments.} We distinguish between CVE-assigned vulnerabilities and acknowledged issues based on maintainer response rather than our own severity assessment. Findings of comparable technical complexity received documentation updates rather than CVEs, reflecting maintainer judgment about exploitability in practice. They were disputed due to concerns about exploitability or limited lateral movement. Since actions were taken, we have respected the maintainers' decisions and have not pursued further escalation, while noting that the vulnerable code patterns remain and that it is up to the individual's security mindset to mitigate them. More precisely, QRS identified the following:

\noindent\textbf{Path Traversal in numpy-2.3.4}~\cite{numpy_github}. NumPy's \texttt{f2py} Fortran Python interface accepts modules that enable attackers to read local files outside intended directories with an include traversal statement. Because exploitation requires the victim to accept and process a malicious \texttt{f90} module, the maintainers raised concerns about exploitability.

\noindent\textbf{Code Execution in numpy-2.3.4}. The \texttt{f2py} Fortran interface may also accept modules that enable attackers to execute arbitrary commands locally, under the same user's privilege context, with a specially crafted Python payload set in a parameter passed down to numpy's \texttt{eval}. Both vulnerabilities were briefly addressed in NumPy's documentation~\cite{numpy_doc}.

\noindent\textbf{DoS in jsonschema-4.25.1}~\cite{jsonschema_github}. The schema validation logic is susceptible to catastrophic backtracking when processing specially crafted malicious JSON schemas in \texttt{pattern} keywords. A commit~\cite{jsonschema_commit} was made by the maintainers regarding the responsibility of users while processing untrusted input data.

\noindent\textbf{Weak RNG \& Race Condition in s3transfer-0.14}~\cite{s3transfer_github}. The S3 download implementation uses a weak random generator for temporary file names that enable a TOCTOU file replacement via symlinks in predicted paths. AWS acknowledged this as informational, citing its Shared Responsibility Model~\cite{aws_shared_responsibility} as the primary mechanism for delegating local filesystem security to customers.

\subsection{The CWE-Bench-Java Dataset}
\label{subsec:cwe-bench-java-set}

To demonstrate language portability and enable direct comparison with IRIS~\cite{iris2025}, we evaluate QRS on CWE-Bench-Java, a benchmark of 120 previously vetted CVEs in real-world Java projects. We create generic \textbf{extra strategy} CWE templates for CWE-022 (Path Traversal: 54 packages), CWE-078 (OS Command Injection: 11 packages), CWE-079 (Cross-Site Scripting: 31 packages), and CWE-094 (Code Injection: 16 packages), and scope our results to these four CWE categories. We exclude 30 packages from this dataset according to \autoref{tab:java-exclusions}, due to unavailability, miscategorization, or build failures of obsolete components.

\begin{table}[ht]\centering
\caption{CWE-Bench-Java: exclusions applied before any QRS execution.}
\label{tab:java-exclusions}\scriptsize
\rowcolors{2}{gray!10}{white}
\begin{tabular}{@{}lcl@{}}\toprule
\textbf{Reason} & \textbf{\#} & \textbf{Determined by} \\ \midrule
Branch or version unavailable upstream  & 13 & Clone attempt \\
Vulnerable code not in Java (JS/XML/JSP) &  6 & File-extension audit \\
Build failure under documented toolchain &  1 & Maven / Gradle build \\
CWE outside the four evaluated classes   & 10 & GitHub Advisory label \\
\midrule \textbf{Total excluded} & \textbf{30} & \\
\textbf{In scope} & \textbf{90} & \\ \bottomrule
\end{tabular}\end{table}

Each project is cloned directly from GitHub at the specific vulnerable branch version and scanned once using the C$_3$ configuration with the appropriate CWE-class preprompt pair for \textbf{Q} and \textbf{R}. In parallel with IRIS' evaluation protocol, QRS also takes a CWE class as an input. The databases are also automatically generated while building these packages from their \texttt{pom.xml} or \texttt{build.gradle} definitions. QRS achieves a 64.4\% CVE detection rate in a single package scan, as shown in \autoref{sec:comparison-llm-research}, compared to IRIS, which achieved 45.8\%. Even in terms of absolute numbers, QRS detected 58 vulnerabilities in a significantly reduced dataset (75\% of the original), while IRIS detected 55.

\subsection{QRS Metrics}
\label{sec:qrs-metrics}

For the Top100 dataset, we run a broad configuration across 1000 scans to obtain robust insights and model behavior using the framework. We collect multi-dimensional performance metrics across QRS's pipeline. The metrics span the triad of agents, including query-generation efficiency, runtime data, LLM classification performance, and tool-calling capabilities, and we compare them with relevant tools and prior research.

\noindent\textbf{Query Generation.} The CodeQL query generation capabilities of each model family vary substantially, as seen in \autoref{tab:codeql-efficiency}, given the absence of pre-training. These discrepancies in configuration outputs have direct implications for the framework's efficiency, underscoring Claude's and Gemini's query-synthesis capabilities relative to the degraded performance of the other models. Lower temperatures, in most cases, seem preferable for deterministic syntactic precision.

\begin{table}[!ht]
\centering
\caption{Top100: CodeQL Agent Efficiency}
\label{tab:codeql-efficiency}
\scriptsize
\rowcolors{2}{gray!10}{white}
\begin{tabular}{@{}lrrrrr@{}}
\toprule
\textbf{Config} & \textbf{\# Queries} & \textbf{W/ Results} & \textbf{W/O Results} & \textbf{Failed} & \textbf{Fail Rate} \\
\midrule
C$_0$ & 1437 & 1007 & 430 & 416 & 28.95\% \\
C$_1$ & 1461 & 993 & 468 & 421 & 28.82\% \\
DC$_0$ & 997 & 621 & 376 & 571 & 57.27\% \\
DC$_1$ & 999 & 579 & 420 & 620 & 62.06\% \\
DR$_0$ & 1516 & 771 & 745 & 861 & 56.79\% \\
DR$_1$ & 1555 & 801 & 754 & 802 & 51.58\% \\
G$_0$ & 1553 & 1089 & 464 & 413 & 26.59\% \\
G$_1$ & 1447 & 930 & 517 & 470 & 32.48\% \\
GPT$_0$ & 1451 & 786 & 665 & 861 & 59.34\% \\
GPT$_1$ & 1371 & 630 & 741 & 935 & 68.20\% \\
\bottomrule
\end{tabular}
\vspace{0.3em}
\begin{flushleft}
\small\textit{\textbf{W/ Results}: queries that compiled and returned findings. \textbf{W/O Results}: queries that compiled but returned no findings. \textbf{Failed}: queries with compilation or validation errors.}
\end{flushleft}
\end{table}

\noindent\textbf{Performance.}
A consistent pattern emerges across all configurations, regarding high accuracy and recall substantially exceeding precision as seen in \autoref{tab:performance-top-100}. This asymmetry is a design decision, as FN (missed vulnerabilities) carries greater risk than FP, since QRS is hunting for unknown vulnerabilities. The trade-off is appropriate for a scanning tool that aims for comprehensive coverage over an analyst's workload, since that workload has also been greatly minimized and contextualized by the \textbf{S} agent.

\begin{table}[!ht]
\centering
\caption{Top100: Prediction Performance Metrics}
\label{tab:performance-top-100}
\scriptsize
\rowcolors{2}{gray!10}{white}
\begin{tabular}{@{}lccccc@{}}
\toprule
\textbf{Config} & \textbf{Correct/Predict} & \textbf{Accuracy} & \textbf{Precision} & \textbf{Recall} & \textbf{F1} \\
\midrule
C$_0$ & 890/952 & 93.49\% & 44.23\% & 92.00\% & 0.597 \\
C$_1$ & 886/942 & 94.06\% & 45.05\% & 87.23\% & 0.594 \\
DC$_0$ & 84/105 & 80.00\% & 22.22\% & 100.00\% & 0.364 \\
DC$_1$ & 72/90 & 80.00\% & 48.57\% & 100.00\% & 0.654 \\
DR$_0$ & 72/131 & 54.96\% & 24.36\% & 100.00\% & 0.392 \\
DR$_1$ & 83/143 & 58.04\% & 24.05\% & 100.00\% & 0.388 \\
G$_0$ & 264/318 & 83.02\% & 37.97\% & 85.71\% & 0.526 \\
G$_1$ & 224/266 & 84.21\% & 43.06\% & 96.88\% & 0.596 \\
GPT$_0$ & 220/283 & 77.74\% & 28.41\% & 100.00\% & 0.442 \\
GPT$_1$ & 163/252 & 64.68\% & 26.27\% & 93.94\% & 0.411 \\
\bottomrule
\end{tabular}
\end{table}

\noindent\textbf{Average configuration metrics.}
Claude and Gemini demonstrate higher token consumption and tend to consume all iterations, as instructed, for a timely and comprehensive scan. GPT consistently did not comply with instructions and terminated prematurely, still yielding results. DeepSeek-chat also tends to finish faster or halt entirely, leading to fewer quality findings for review, compared to DeepSeek-reasoner, which performs in a more decent and cost-effective manner, as seen in \autoref{tab:avg-metrics-top-100}. LLM API call caching from providers is not taken into account, as we cannot measure this effectively, so the costs refer to the worst case scenario (no-caching)~\cite{claudecaching, deepseekcaching, geminicaching, gptcaching}. Totals computed from raw logs are rounded.

\begin{table}[!ht]
\centering
\caption{Top100: Average Performance Metrics}
\label{tab:avg-metrics-top-100}
\scriptsize
\rowcolors{2}{gray!10}{white}
\begin{tabular}{@{}llcccc@{}}
\toprule
\textbf{Config} & \textbf{Avg Runtime} & \textbf{Avg Cost} & \textbf{Avg Tokens} & \textbf{Q + R + S Iter.} \\
\midrule
C$_0$ & 12m22s & \$1.453 & 706k & 19 + 18 + 12\\
C$_1$ & 9m48s & \$1.228 & 561k & 19 + 16 + 12 \\
DC$_0$ & 10m1s & \$0.061 & 212k & 18 + 6 + 3\\
DC$_1$ & 9m57s & \$0.064 & 221k & 18 + 5 + 3 \\
DR$_0$ & 21m26s & \$0.078 & 263k & 19 + 10 + 3 \\
DR$_1$ & 19m17s & \$0.080 & 268k & 19 + 11 + 3 \\
G$_0$ & 11m42s & \$1.034 & 406k & 20 + 9 + 4 \\
G$_1$ & 11m19s & \$0.989 & 383k & 20 + 8 + 3 \\
GPT$_0$ & 8m7s & \$0.2240 & 153k & 12 + 4 + 5 \\
GPT$_1$ & 8m12s & \$0.2232 & 152k & 12 + 5 + 5 \\
\midrule
Total & 203h 45m & \$543.34 & 332.5M & 30646 \\
\bottomrule
\end{tabular}
\end{table}

\noindent\textbf{Toolset calling.}
As instructed, all \textbf{Q} agents run package profiling only once and then start executing their queries. Gemini tends to avoid fixing queries and instead relies on brute-force direct execution. Claude uses the toolsets more consistently, and achieved the broadest coverage. All configurations seem to prefer extracting raw code snippets, and most avoid directly tracing code flows (flows may also be provided by the \textbf{Q} agent, diminishing this need). API calls include parallel function calling~\cite{claudemultitools, geminimultitools, gptmultitools, deepseekmultitools}, showing inflated tool call summaries versus the maximum iteration setting \autoref{tab:tool-calls-top100}.

\begin{table}[!ht]
\centering
\caption{Top 100: Average Tool Calling}
\label{tab:tool-calls-top100}
\scriptsize
\rowcolors{2}{gray!10}{white}
\setlength{\tabcolsep}{3pt}
\begin{tabular}{@{}l|ccc|cccc|c@{}}
\toprule
& \multicolumn{3}{c|}{CodeQL Tools} & \multicolumn{4}{c|}{Reviewer Tools} & Sanity \\
\cmidrule(lr){2-4} \cmidrule(lr){5-8} \cmidrule(lr){9-9}
Config & pp & eq & fq & ecs & gs & tf & sf & Iterations \\
\midrule
C$_0$ & 1.0 & 12.7 & 1.9 & 27.1 & 3.5 & 1.8 & 12.3 & 12 \\
C$_1$ & 1.0 & 12.7 & 2.0 & 25.3 & 3.3 & 1.6 & 11.5 & 12 \\
DC$_0$ & 1.0 & 8.6 & 2.2 & 3.9 & 1.0 & 0.0 & 1.1 & 3 \\
DC$_1$ & 1.0 & 8.3 & 2.7 & 3.0 & 1.0 & 0.0 & 0.9 & 3 \\
DR$_0$ & 1.0 & 13.1 & 2.9 & 7.1 & 2.0 & 0.0 & 1.5 & 3 \\
DR$_1$ & 1.0 & 12.9 & 3.4 & 7.6 & 2.1 & 0.0 & 1.6 & 3 \\
G$_0$ & 1.0 & 14.9 & 0.7 & 5.3 & 0.9 & 0.0 & 3.5 & 4 \\
G$_1$ & 1.0 & 13.6 & 0.9 & 4.9 & 0.7 & 0.0 & 3.0 & 3 \\
GPT$_0$ & 1.0 & 12.5 & 3.2 & 6.8 & 0.6 & 0.0 & 3.3 & 5 \\
GPT$_1$ & 1.0 & 11.7 & 3.1 & 5.7 & 0.7 & 0.1 & 3.0 & 5 \\
\bottomrule
\end{tabular}
\vspace{0.5em}
\begin{center}
\small\textit{\textbf{Q Max Iterations}: 20, \textbf{R Max Iterations}: 10 $\times$ N (N: number of finding clusters). \textbf{Notation:} pp: package\_profile, eq: execute\_query, fq: fixup\_query, ecs: extr\_code\_snip, gs: grep\_search, tf: trace\_flow, sf: submit\_finding}
\end{center}
\end{table}

Based on the nature of the Top100 packages, \textbf{Q} hypotheses and \textbf{R} findings spanned across 115 CWEs, with the most important categories being:
\begin{enumerate*}[label=(\roman*)]
    \item \textit{Injection \& Code Execution}: CWE-502, CWE-094, CWE-078 with a category subtotal of 1369 findings,
    \item \textit{Path Traversal \& File System Access}: CWE-022, CWE-073, CWE-059 with a subtotal of 670,
    \item \textit{Denial-of-Service \& Resource Exhaustion}: CWE-1333, CWE-400 with a subtotal of 578,
    \item \textit{Cryptographic Weaknesses}: CWE-798, CWE-327 with a subtotal of 470, and
    \item \textit{Race Conditions \& Concurrency}: CWE-362, CWE-367 with a subtotal of 87
\end{enumerate*}, with Claude dominating across all categories.

\noindent\textbf{False Positive reduction.}
QRS addresses alert fatigue by utilizing the \textbf{S} agent as a post-review filtering hook. In our highest-volume configuration (C$_0$) with a 64.8\% reduction rate, the security expert would have to review 140 TP findings for the Top100 dataset. Claude also tends to dominate total finding counts, but reduces final verdicts to manageable numbers as seen in \autoref{tab:r-to-s-reduction}.

\begin{table}[ht]
\centering
\caption{Top 100: TP Finding Reduction from R to S Agent}
\label{tab:r-to-s-reduction}
\scriptsize
\rowcolors{2}{gray!10}{white}
\begin{tabular}{@{}lccc@{}}
\toprule
\textbf{Config} & \textbf{R $\rightarrow$ S} & \textbf{Reduction} & \textbf{Reduction} \% \\
\midrule
C$_0$ & 398 $\rightarrow$ 140 & 258 & 64.8\% \\
C$_1$ & 300 $\rightarrow$ 103 & 197 & 65.7\% \\
DC$_0$ & 73 $\rightarrow$ 27 & 46 & 63.0\% \\
DC$_1$ & 61 $\rightarrow$ 35 & 26 & 42.6\% \\
DR$_0$ & 99 $\rightarrow$ 90 & 9 & 9.1\% \\
DR$_1$ & 104 $\rightarrow$ 85 & 19 & 18.3\% \\
G$_0$ & 104 $\rightarrow$ 85 & 19 & 18.3\% \\
G$_1$ & 87 $\rightarrow$ 80 & 7 & 8.0\% \\
GPT$_0$ & 173 $\rightarrow$ 104 & 69 & 39.9\% \\
GPT$_1$ & 203 $\rightarrow$ 129 & 74 & 36.5\% \\
\bottomrule
\end{tabular}
\vspace{0.5em}
\begin{center}
\small (a) The G$_0$ and DR$_1$ rows are genuinely identical. (b) Denominator note vs ablation: Counts cover all submissions across scans. \autoref{tab:ablation-stages} restricts to the subset of scans with a human-reviewed ground-truth universe; therefore, it reports smaller absolute counts.
\end{center}
\end{table}

\noindent\textbf{False Negative rate.} The framework's primary limitation is precision, as the FN rate is minimal and gets eliminated with overlapping scans. More precisely, the FN rate in Top100 is just 0.52\% (18 FN). We stress that this is not codebase-level recall, which cannot be measured over a target whose full vulnerability set is unknown. It is the rate at which this stage preserves a finding the pipeline had already surfaced and a human confirmed. Codebase-level recall can only be reported for Hist20 and CWE-Bench-Java.

\subsection{Ablation Study}
\label{sec:ablation}

A recurring question in pipelined multi-agent design is whether each stage incurs its own cost or benefits the output, and whether a single agent with the union of toolsets could perform comparably. Across the Top100 corpus four ablations follow: \begin{enumerate*}[label=(\roman*)]
    \item stage-wise detection quality,
    \item code-reduction effect from \textbf{Q},
    \item tool preference evidence in a single agent architecture, and
    \item per-agent, per-tool resource decomposition.
\end{enumerate*}

\subsubsection{Stage-wise Ablation (Q / Q+R / Q+R+S)}

\autoref{tab:ablation-stages} evaluates the pipeline after each stage against the human-evaluated ground-truth universe. The \textbf{Q} stage is all-positive by construction, as CodeQL emits findings without a verdict; we therefore characterize it by the size of the alert universe $n$ and by the fraction $\pi_Q$ of those alerts a human ultimately confirmed.

\begin{table*}[t]
\centering
\caption{Stage-wise ablation over the full Top100 corpus (1{,}000 scans).
$n$ is the deduplicated ground-truth universe and $n^{+}$ its positive subset;
$\pi_Q=n^{+}/n$ is the precision of the raw \textbf{Q} alert stream. F1$_\mu$ is pooled over findings; F1$_M$ is the mean of
per-scan F1. TN is omitted and recoverable as $n-\mathrm{TP}-\mathrm{FP}-\mathrm{FN}$.}
\label{tab:ablation-stages}
\scriptsize
\setlength{\tabcolsep}{2.6pt}
\rowcolors{3}{gray!10}{white}
\begin{tabular}{@{}lrrc|rrrcccc|rrrcccc@{}}
\toprule
& \multicolumn{3}{c|}{\textbf{Q} (raw alerts)}
& \multicolumn{7}{c|}{\textbf{Q+R}}
& \multicolumn{7}{c}{\textbf{Q+R+S}} \\
\cmidrule(lr){2-4}\cmidrule(lr){5-11}\cmidrule(lr){12-18}
\textbf{Config} & $n$ & $n^{+}$ & $\pi_Q$
& TP & FP & FN & Acc & Prec & Rec & F1$_\mu$
& TP & FP & FN & Acc & Prec & Rec & F1$_\mu$ \\
\midrule
C$_0$   &  952 &  50 & 0.053 &  50 &  264 & 0 & 0.723 & 0.159 & 1.000 & 0.275 &  46 &  58 &  4 & 0.935 & 0.442 & 0.920 & 0.597 \\
C$_1$   &  942 &  47 & 0.050 &  46 &  234 & 1 & 0.751 & 0.164 & 0.979 & 0.281 &  41 &  50 &  6 & 0.941 & 0.451 & 0.872 & 0.594 \\
DC$_0$  &  105 &   6 & 0.057 &   6 &   67 & 0 & 0.362 & 0.082 & 1.000 & 0.152 &   6 &  21 &  0 & 0.800 & 0.222 & 1.000 & 0.364 \\
DC$_1$  &   90 &  17 & 0.189 &  17 &   43 & 0 & 0.522 & 0.283 & 1.000 & 0.442 &  17 &  18 &  0 & 0.800 & 0.486 & 1.000 & 0.654 \\
DR$_0$  &  131 &  19 & 0.145 &  19 &   68 & 0 & 0.481 & 0.218 & 1.000 & 0.358 &  19 &  59 &  0 & 0.550 & 0.244 & 1.000 & 0.392 \\
DR$_1$  &  143 &  19 & 0.133 &  19 &   77 & 0 & 0.462 & 0.198 & 1.000 & 0.330 &  19 &  60 &  0 & 0.580 & 0.241 & 1.000 & 0.388 \\
G$_0$   &  318 &  35 & 0.110 &  33 &   61 & 2 & 0.802 & 0.351 & 0.943 & 0.512 &  30 &  49 &  5 & 0.830 & 0.380 & 0.857 & 0.526 \\
G$_1$   &  266 &  32 & 0.120 &  31 &   48 & 1 & 0.816 & 0.392 & 0.969 & 0.559 &  31 &  41 &  1 & 0.842 & 0.431 & 0.969 & 0.596 \\
GPT$_0$ &  283 &  25 & 0.088 &  25 &  127 & 0 & 0.551 & 0.164 & 1.000 & 0.282 &  25 &  63 &  0 & 0.777 & 0.284 & 1.000 & 0.442 \\
GPT$_1$ &  252 &  33 & 0.131 &  32 &  154 & 1 & 0.385 & 0.172 & 0.970 & 0.292 &  31 &  87 &  2 & 0.647 & 0.263 & 0.939 & 0.411 \\
\midrule
\textbf{All} & \textbf{3482} & \textbf{283} & \textbf{0.081}
& \textbf{278} & \textbf{1143} & \textbf{5} & \textbf{0.670} & \textbf{0.196} & \textbf{0.982} & \textbf{0.326}
& \textbf{265} & \textbf{506}  & \textbf{18} & \textbf{0.850} & \textbf{0.344} & \textbf{0.936} & \textbf{0.503} \\
\bottomrule
\end{tabular}
\vspace{0.25em}
\begin{flushleft}
\small\textit{Macro-averaged F1 (F1$_M$), aggregate: 0.172 at \textbf{Q+R} and 0.192 at \textbf{Q+R+S}. The gap against F1$_\mu$ reflects the estimator, not the pipeline: per-scan positive counts are small, so pooling favors high-yield scans. We report both wherever we compare against prior work. Recall is the retention rate of findings.}
\end{flushleft}
\end{table*}

Three results follow. First, the symbolic stage alone is not deployable: only $8.1\%$ of raw \textbf{Q} alerts survive expert review, comparable to the raw alert streams of the rule-based tools in \autoref{sec:observations}. Query synthesis localizes the attack surface; it does not verify it. Second, \textbf{R} recovers almost the entire positive set, $278/283$ ($98.2\%$), but
at $19.6\%$ precision: it is a high-sensitivity, low-specificity filter by design, since a missed vulnerability is costlier than a reviewed false alarm. Third, \textbf{S} converts that stream into a workable queue. It discards $637$ of \textbf{R}'s $1{,}143$ false positives ($55.7\%$) while surrendering $13$ of its $278$ true positives ($4.7\%$), raising precision by $75.5\%$ relative and pooled F1 by $54.3\%$ relative. End to end, the triad reduces a queue of $3{,}482$ raw alerts to $771$ reviewed findings, a $77.9\%$ reduction, while preserving $93.6\%$ of the confirmed positives.
 
The $4.7\%$ retention loss at \textbf{S} is a real cost. It is concentrated in four configurations (C$_0$: 4, C$_1$: 5, G$_0$: 3, GPT$_1$: 1), while both DeepSeek variants, G$_1$ and GPT$_0$ discard no confirmed positive at all. Because \textbf{S} is detachable, an operator unwilling to accept any retention loss can consume \textbf{R}'s artifact directly at $19.6\%$ precision.

\subsubsection{Pipelined Code Reduction}

A central architectural claim is that \textbf{Q} acts as a slicer, allowing \textbf{R} to reason over a small fraction of a package rather than ingesting it wholesale. \autoref{tab:ablation-reduction} quantifies this by comparing the unique source lines \textbf{R} actually reads through the raw code reading tool against each package's total line count.

\begin{table}[!ht]
\centering
\caption{Code reduction: source actually read by \textbf{R} versus total
package size, pooled over 1{,}000 scans. \emph{Uniq} merges overlapping
ranges. Macro weights every scan equally; micro is pooled and therefore
weighted by package size.}
\label{tab:ablation-reduction}
\scriptsize
\setlength{\tabcolsep}{3pt}
\rowcolors{3}{gray!10}{white}
\begin{tabular}{@{}lrrrrrr@{}}
\toprule
& & \multicolumn{2}{c}{\textbf{Lines}} & \multicolumn{2}{c}{\textbf{Reduction}} & \\
\cmidrule(lr){3-4}\cmidrule(lr){5-6}
\textbf{Stratum} & \textbf{Scans} & Total & Uniq & macro & micro & \textbf{Files Touched} \\
\midrule
\multicolumn{7}{@{}l}{\textit{By package size}} \\
$<$1K            &  240 &    490 & 101 & 70.97\% & 79.44\% & 28.23\% \\
1K--5K           &  310 &   2978 & 156 & 93.61\% & 94.77\% & 10.67\% \\
5K--20K          &  310 &  10356 & 163 & 98.23\% & 98.43\% &  6.53\% \\
20K--100K        &   90 &  39331 & 223 & 99.35\% & 99.43\% &  2.63\% \\
$>$100K          &   50 & 175083 & 261 & 99.85\% & 99.85\% &  1.36\% \\
\midrule
\textbf{Stratum} & \textbf{Scans} & Avg. Lines & Uniq & macro & micro & \textbf{Files Touched} \\
\midrule
\multicolumn{7}{@{}l}{\textit{By model (temperatures merged)}} \\
C$_0,_1$            &  200 & 16545 & 417 & 77.92\% & 97.48\% & 8.50\% \\
DC$_0,_1$           &  200 & 16545 &  59 & 95.95\% & 99.64\% & 2.51\% \\
DR$_0,_1$           &  200 & 16545 & 101 & 93.49\% & 99.39\% & 3.55\% \\
G$_0,_1$            &  200 & 16545 & 139 & 88.71\% & 99.16\% & 4.43\% \\
GPT$_0,_1$          &  200 & 16545 &  65 & 96.12\% & 99.61\% & 3.57\% \\
\midrule
\textbf{All}      & \textbf{1000} & \textbf{16545} & \textbf{156}
                  & \textbf{90.44\%} & \textbf{99.06\%} & \textbf{4.51\%} \\
\bottomrule
\end{tabular}
\vspace{0.25em}
\begin{flushleft}
\small\textit{Package sizes span 46--245{,}973 lines. \textbf{R} issues a mean
of 13.73 \texttt{extract\_code\_snippet} calls per scan, touching 2.53 of 56
files.}
\end{flushleft}
\end{table}

Across the corpus, \textbf{R} covers a mean of $156$ unique lines across $2.53$ files, against a mean package size of $16{,}545$ lines across 56 files: $4.5\%$ of files and $9.6\%$ of lines under macro averaging. The decisive property is that reduction \emph{grows} with codebase size, from $71.0\%$ on packages below 1K lines to $99.9\%$ above 100K lines, because the number of
locations CodeQL flags is governed by the density of candidate patterns rather
than by total volume. This is the mechanism behind the scaling behavior reported in \autoref{tab:avg-metrics-top-100}: the review stage's context requirement is approximately invariant to package size, which is what distinguishes query-first triage from whole-file or whole-repository prompting, whose context cost is linear in the codebase. Claude displays the lowest reduction rate because it inspects roughly 7 times more unique code than the cheapest configuration, due to better querying capabilities, while yielding better results and more actual findings.
 
We note the direct lineage of this objective. Yamaguchi et al.~\cite{yamaguchi2015} reported a $94.9\%$ reduction in code requiring manual inspection using automatically inferred CPG traversals, the same goal pursued a decade earlier without language models. QRS obtains a comparable macro-averaged reduction ($90.4\%$) from synthesized CodeQL queries
rather than clustered definition--use patterns, and additionally reasons with the surviving locations rather than returning them for plain manual audit.

\subsubsection{Evidence bearing on a single-agent design}

We are careful not to claim that a single agent holding the union of the three toolsets is architecturally impossible; it is not in recent 1M context models, but looks impossible with 200k ones. The claim we defend is narrower and empirically grounded: at the iteration and context budgets used here, a merged agent is dominated. Two measurements support this.

\noindent\textbf{Budget exhaustion.} \textbf{Q} is the most narrowly scoped agent in the pipeline as it may only author, execute and repair CodeQL queries and it still exhausts the context budget with a 20-iteration session in 38.5\% of total scans. \textbf{R} receives a fresh budget per finding cluster rather than one shared budget for the session and exhausts it only $7.9\%$ out of the entire corpus. A merged agent would face the union of these demands inside a single context window and as seen in \autoref{tab:avg-metrics-top-100}, the pipeline already exceeds the 200K window of the models used, before accounting for the compaction losses that consolidation would require.

\noindent\textbf{Tool preference.} \textbf{R} is the only agent offered both a raw code-reading toolset, a grep search, and a structured, query-backed dataflow tracer for the same task of justifying a finding. Its first action is raw code reading in $1{,}637$ of $1{,}654$ review groups ($99.0\%$). Raw reads outnumber structured traces $11{,}222$ to $351$ ($32.0{:}1$), and in 5/10 configurations the structured tool is never invoked at all. This holds across all model families at both temperatures, despite explicit prompt instructions. The architectural consequence is that \textbf{Q}'s tool palette is deliberately restricted to CodeQL operations: not because a merged agent cannot be built, but because it reliably abandons the symbolic component that produces the code reduction of \autoref{tab:ablation-reduction}. We note that this is observational evidence from a pipeline designed around the separation, not a controlled merged-agent re-run under an equalized budget, which we identify as future work.

\begin{table}[!ht]
\centering
\caption{Budget exhaustion and tool preference under free choice.
\emph{Q cap}: scans in which \textbf{Q} exhausted its 20-iteration session
budget. \emph{R cap}: finding-groups in which \textbf{R} exhausted its
per-group budget. \emph{R first}: groups whose first \textbf{R} tool call was
raw code reading rather than structured flow tracing.
\emph{raw:str}: overall call-volume ratio of raw reads to structured traces.}
\label{tab:ablation-load}
\scriptsize
\setlength{\tabcolsep}{4pt}
\rowcolors{2}{gray!10}{white}
\begin{tabular}{@{}lrrrr@{}}
\toprule
\textbf{Config} & \textbf{Q cap} & \textbf{R cap} & \textbf{R first = raw} & \textbf{raw:str} \\
\midrule
C$_0$   & 29.0\% &  5.9\% & \phantom{0}99.2\% (251/253) & 17.2 : 1 \\
C$_1$   & 37.0\% &  6.1\% & 100.0\% (230/230)           & 17.5 : 1 \\
DC$_0$  & 33.0\% & 16.2\% & 100.0\% (105/105)           & 486 : 0  \\
DC$_1$  & 40.0\% &  8.2\% & 100.0\% \phantom{0}(98/98)  & 403 : 0  \\
DR$_0$  & 40.4\% & 31.3\% & 100.0\% (115/115)           & 453.5 : 1 \\
DR$_1$  & 41.0\% & 30.4\% & \phantom{0}99.2\% (124/125) & 965 : 0  \\
G$_0$   & 80.0\% &  0.0\% & \phantom{0}97.3\% (251/258) & 621 : 0  \\
G$_1$   & 84.0\% &  0.4\% & \phantom{0}97.8\% (221/226) & 561 : 0  \\
GPT$_0$ &  0.0\% &  0.0\% & 100.0\% (117/117)           & 753 : 0  \\
GPT$_1$ &  0.0\% &  0.8\% & \phantom{0}98.4\% (125/127) & 70.6 : 1 \\
\midrule
\textbf{All} & \textbf{38.5\%} & \textbf{7.9\%}
             & \textbf{99.0\% (1637/1654)} & \textbf{32.0 : 1} \\
\bottomrule
\end{tabular}
\vspace{0.25em}
\begin{flushleft}
\small\textit{Aggregate denominators: \textbf{Q} 1{,}000 scans, 1{,}654 finding-groups, 11{,}222 raw calls versus 351 structured calls. The structured tool was the \emph{first} action in \textbf{0} of 1{,}654 groups.}
\end{flushleft}
\end{table}

\subsubsection{Per-agent resource decomposition}

\autoref{tab:ablation-cost} decomposes token, cost and wall-clock spend per scan across the three agents and their individual tools, pooled over all 1{,}000 scans. \textbf{Q} dominates the budget, accounting for $63\%$ of tokens and $76.6\%$ of cost, driven by the iterative execute-and-refine loop and by recovery from the CodeQL compilation failures reported in \autoref{tab:codeql-efficiency}, with \texttt{execute\_query} and \texttt{fixup\_query} alone consuming in average $\$0.297$ of the $\$0.543$ per-scan total. \textbf{Q} also owns $85.7\%$ of subprocess wall-clock, since CodeQL evaluation is the only genuinely compute-bound operation in the pipeline.
 
The corollary is the direct answer to whether the two downstream agents earn their place. \textbf{R} and \textbf{S} together consume $23.4\%$ of pipeline spend and $37\%$ of tokens, and it is that $23.4\%$ that carries precision from $8.1\%$ to $34.4\%$, a $4.2\times$ improvement, while retaining $93.6\%$ of confirmed positives. \textbf{S} in particular is the cheapest stage in the pipeline at $5.5\%$ of cost, and removes $55.7\%$ of the false positives
reaching it.

\begin{table}[!ht]
\centering
\caption{Per-agent, per-tool resource decomposition; mean per scan over
1{,}000 scans. \emph{Exec} is subprocess wall-clock (CodeQL compilation and
evaluation, CVE retrieval); in-process tools are sub-second.
\emph{rsn} is model reasoning not attributable to a tool call.}
\label{tab:ablation-cost}
\scriptsize
\setlength{\tabcolsep}{4pt}
\rowcolors{2}{gray!10}{white}
\begin{tabular}{@{}llrrrr@{}}
\toprule
\textbf{Agent} & \textbf{Tool} & \textbf{Calls} & \textbf{Tokens} & \textbf{Cost \$} & \textbf{Exec (s)} \\
\midrule
\multirow{6}{*}{\textbf{Q}}
{\textbf{Q}} & \texttt{pp}  &  1.00 &   3888 & 0.0061 &   0.19 \\
 & \texttt{eq}  & 12.03 & 116180 & 0.2498 & 232.55 \\
 & \texttt{fq}  &  2.29 &  30505 & 0.0471 & 119.81 \\
 & \texttt{fr}  &  1.00 &  15418 & 0.0331 &   0.00 \\
 & \texttt{rsn} &  ---  &  43458 & 0.0799 &   0.00 \\
\cmidrule(l){2-6}
 & \textbf{total} & \textbf{16.32} & \textbf{209449} & \textbf{0.4161} & \textbf{352.55} \\
\midrule
\multirow{6}{*}{\textbf{R}}
{\textbf{R}} & \texttt{ecs} &  9.68 &  48423 & 0.0418 & 0.00 \\
 & \texttt{gs}  &  1.56 &   9957 & 0.0073 & 0.00 \\
 & \texttt{tf}  &  0.35 &   2611 & 0.0023 & 9.72 \\
 & \texttt{sf}  &  4.16 &  27029 & 0.0318 & 0.00 \\
 & \texttt{cr}  &  0.88 &   8182 & 0.0078 & 0.00 \\
 & \texttt{rsn} &  ---  &   5913 & 0.0060 & 0.00 \\
\cmidrule(l){2-6}
 & \textbf{total} & \textbf{16.63} & \textbf{102115} & \textbf{0.0970} & \textbf{9.72} \\
\midrule
\multirow{6}{*}\_
{\textbf{S}} & \texttt{vf} & 4.06 & 20842 & 0.0298 & 48.95 \\
\cmidrule(l){2-6}
 & \textbf{total} & \textbf{4.06} & \textbf{20842} & \textbf{0.0298} & \textbf{48.95} \\
\midrule
\multicolumn{2}{@{}l}{\textbf{Pipeline}} & \textbf{37.01} & \textbf{332406} & \textbf{0.5429} & \textbf{411.22} \\
\multicolumn{2}{@{}l}{\emph{share of total}} & --- & \emph{63/31/6\%} & \emph{77/18/5\%} & --- \\
\bottomrule
\end{tabular}
\vspace{0.25em}
\begin{flushleft}
\small\textit{Notation: \texttt{pp}: \texttt{package\_profile}, \texttt{eq}: \texttt{execute\_query}, \texttt{fq}: \texttt{fixup\_query}, \texttt{fr}: \texttt{finalize\_report}, \texttt{ecs}: \texttt{extract\_code\_snippet}, \texttt{gs}: \texttt{grep\_search}, \texttt{tf}: \texttt{trace\_flow}, \texttt{sf}: \texttt{submit\_finding}, \texttt{cr}: \texttt{complete\_review}, \texttt{vf}: \texttt{verify\_finding} (composite of \texttt{search\_cve} and the contextual verification call), \texttt{rsn}: \texttt{model reasoning}. Costs assume no provider-side prompt caching and are therefore an upper bound. Tools such as \texttt{fr} or \texttt{sf} do not consume measurable execution time leading to near-to-zero or zero captured values.}
\end{flushleft}
\end{table}

\subsection{CWE-Bench-Java}

For this dataset, we use a single configuration rather than the 10-configuration sweep applied to earlier Python datasets. Additionally, the preprompt approach requires curation of 4 CWE templates to properly steer the two agents. As the in-scope packages from this dataset may reach up to 1.25GB of raw code size, QRS is able to handle them relatively easily, as \textbf{Q} is able to slice the code to points-of-interest, so that \textbf{R} can then focus on minimal lines of code. As we use the latest Claude Sonnet 4.6 model across the triad of agents and scan huge Java codebases versus the max 65MB Python ones, we see a proportionate inflation in costs, tokens, and runtimes. See \autoref{tab:cwbench-java} for overall metrics.

\begin{table}[t]
\centering
\caption{CWE-Bench-Java: QRS evaluation across 4 CWE categories (C$_3$ configuration)}
\label{tab:cwbench-java}
\scriptsize

\rowcolors{2}{gray!10}{white}
\begin{subtable}[t]{0.49\columnwidth}
\begin{tabular}{@{}lr@{}}
\toprule
\multicolumn{2}{@{}l}{\textbf{CVE Detection}} \\
\midrule
Projects (in-scope / total)     & 90 / 120 \\
Detected                        & 58 / 90 \\
Detection Rate                  & 64.44\% \\
\quad vs.\ IRIS & 55 / 120 \\
\midrule
\multicolumn{2}{@{}l}{\textbf{Classification Performance}} \\
\midrule
Predictions                     & 1097 \\
Accuracy                        & 87.60\% \\
Precision                       & 75.70\% \\
Recall                          & 98.79\% \\
F1 Score                        & 0.857 \\
\bottomrule
\end{tabular}
\end{subtable}
\begin{subtable}[t]{0.49\columnwidth}
\scriptsize
\rowcolors{2}{gray!10}{white}
\begin{tabular}{@{}lr@{}}
\toprule
\multicolumn{2}{@{}l}{\textbf{CodeQL Agent Efficiency}} \\
\midrule
Queries Generated               & 1944 \\
\quad With Results               & 1530 (78.7\%) \\
\quad Failed                     & 412 (21.2\%) \\
\toprule
\multicolumn{2}{@{}l}{\textbf{Operational Cost}} \\
\midrule
Total Runtime                   & 42h 22m \\
Avg.\ Runtime / Project         & 29m 54s \\
Total Cost (USD)                & \$1059 \\
Avg.\ Cost / Project            & \$12.46 \\
Avg.\ Tokens / Project          & 4.0M \\
Avg.\ Iterations (Q / R / S)    & 15.7 / 37.2 / 13.4 \\
\bottomrule
\end{tabular}
\end{subtable}
\end{table}

\section{Observations}
\label{sec:observations}

\subsection{Comparison against SAST tools}

To demonstrate the advantages over purely syntactic pattern matching, we benchmark QRS against 3 widely-known open-source tools, namely Opengrep-v.1.15.1~\cite{opengrep}, an open-source fork of Semgrep, using the standard community Python Semgrep ruleset~\cite{semgrep_rules}, Bandit-v1.9.3~\cite{bandit}, a more sophisticated AST-based Python scanner, and CodeQL v2.23.0 using the library's ruleset~\cite{codeqlruleset2025}.

The former tools exhibited severe alert fatigue on the Top100 dataset. More precisely, Opengrep generated 1251 warnings, Bandit generated 653 medium-to-high-severity alerts (more than 48k if low-severity alerts are included), and CodeQL generated 136 (none of its locations by \texttt{file:line} matched the actual vulnerability location). Compared with our noisiest configuration (i.e., C$_0$), which produced 140 TP candidate findings, we demonstrate a better signal-to-noise ratio than these tools and generate actual, meaningful findings. QRS was also able to trace entire flows and generate the entire vulnerability context available to a security reviewer, fast-forwarding the manual process to only a set of findings that resulted in multiple CVE attributions or maintainer acknowledgments.

Manual sampling of the highest severity alerts confirmed that only rule-based pattern detections (e.g., \texttt{eval} or \texttt{pickle.load}) overlap with QRS's findings, rapidly increasing a security reviewer's workload as multiple findings were not externally controlled or passed through a mitigator or sanitizer, effectively deeming the finding as a FP. QRS traced the code flow and reduced the sheer volume of FPs.

Despite being actively maintained, packages hosted on GitHub are scanned by CodeQL with GitHub's Advanced Security~\cite{githubadvancedsecurity}, Dependabot~\cite{dependabot}, and even AI Copilot~\cite{copilot_autofix} as inferred from repository "Actions" tab~\cite{filelockgithubaction, setuptoolsgithubaction, virtualenvgithubaction}; these vulnerabilities were entirely missed. QRS effectively used CodeQL queries, natively supported by GitHub, to scan the repositories, and none of these vulnerabilities were previously detected. This demonstrates that LLM-driven query synthesis can pave the way for generating detection logic that anticipates patterns not yet codified.

\subsection{Comparison against LLM-based Research}
\label{sec:comparison-llm-research}

\noindent\textbf{Same dataset comparison.} Direct metric comparison across LLM-based vulnerability tools remains difficult because datasets, input requirements, and evaluation granularity all differ. We address this by evaluating QRS on CWE-Bench-Java, which enables a same-dataset comparison with IRIS~\cite{iris2025} and QLCoder~\cite{qlcoder2025}. Prior work does not report the full set of classification metrics, so we compare inputs, CVE detection rates, and F1 scores. Note that prior work does not report all LLM classification metrics as shown in \autoref{tab:framework-comparison}, compared to QRS, and as such, we compare inputs, CVE detection rates, and F1 scores. Estimator choice matters here, and we make it explicit. IRIS reports the mean of per-project F1; a figure pooled over all submitted findings is systematically higher when per-project positive counts are small. Our own Top100 corpus yields $\text{F1}_\mu = 0.503$ against $\text{F1}_M = 0.192$. We therefore report both for QRS and compare only like estimators.

\begin{table}[!ht]
\centering
\caption{Comparison on CWE-Bench-Java. $\text{F1}_\mu$ is pooled over all
submitted findings; $\text{F1}_M$ is the mean of per-project F1, which is the
estimator IRIS reports. Only the $\text{F1}_M$ column is directly comparable
across rows. Detection rates are computed over different denominators; see
\autoref{subsec:cwe-bench-java-set} for the bounded comparison.}
\label{tab:framework-comparison}
\scriptsize
\setlength{\tabcolsep}{3.5pt}
\rowcolors{2}{gray!10}{white}
\begin{tabular}{@{}lccccccc@{}}
\toprule
\textbf{Framework} & \textbf{Input} & \textbf{Det.\ Rate} & \textbf{Acc.} &
\textbf{Prec.} & \textbf{Rec.} & \textbf{F1$_\mu$} & \textbf{F1$_M$} \\
\midrule
CodeQL  & built-in rules & 22.50\%$^\dagger$ & --- & --- & --- & --- & 0.076 \\
IRIS    & CWE spec       & 45.83\%$^\dagger$ & --- & --- & --- & --- & 0.177 \\
QLCoder & CVE meta+diff  & 53.40\%$^\dagger$ & --- & --- & --- & 0.700$^\S$ & --- \\
\midrule
QRS C\textsubscript{3} & CWE spec       & \textbf{64.44\%}$^\ddagger$ &
\textbf{87.6\%} & \textbf{75.7\%} & \textbf{98.8\%} & \textbf{0.857} & \textbf{0.794} \\
\bottomrule
\end{tabular}
\vspace{0.3em}
\begin{flushleft}
\small$^\dagger$over all 120 projects.\quad$^\ddagger$over our 90 in-scope projects.\quad$^\S$QLCoder does not state which estimator it uses; we reproduce the published figure and exclude it from the $\text{F1}_M$ column.
\end{flushleft}
\end{table}

Detection rates are likewise not directly comparable, since ours is computed over 90 in-scope projects and IRIS's over all 120. We bound rather than assume this comparison in \autoref{subsec:cwe-bench-java-set}: under any possible allocation of IRIS's 55 detections across the excluded projects, its rate on our subset cannot exceed $61.1\%$, which is below QRS's $64.4\%$.

QLCoder and CVE-Genie~\cite{cvegenie2025} require known vulnerability metadata, and reproduction systems benefit from ground truth that discovery systems lack: QLCoder takes the CVE description and fix diff as input. QRS and IRIS operate without training data or vulnerability metadata, which is what makes the comparison in \autoref{tab:framework-comparison} meaningful despite the denominator difference.

The precision gap between our two settings is informative in itself. On CWE-Bench-Java, where \textbf{Q} and \textbf{R} receive a CWE-class preprompt and the ground truth is externally labeled, QRS reaches $75.7\%$ precision. On the untargeted Top100 corpus the best configuration reaches $48.6\%$ (DC$_1$, \autoref{tab:performance-top-100}). This is a property of the task rather than of the framework: targeted scanning against a labeled benchmark is a materially easier problem than open-world discovery, and metrics from the two settings should not be pooled.

\noindent\textbf{Cross-dataset limitations.} Metrics from pattern-based classifiers (VulSim~\cite{vulsim2024}, VulBERTa~\cite{hanif2022vulberta}, ReGVD~\cite{nguyen2022regvdrevisitinggraphneural}) and from multi-agent frameworks (LLM4Vuln~\cite{sun2025llm4vulnunifiedevaluationframework}, VulAgent~\cite{wang2025vulagenthypothesisvalidationbasedmultiagent}) are obtained on curated function-level benchmarks such as SVEN and PrimeVul, which differ fundamentally from QRS's whole-package evaluation. These approaches detect vulnerabilities that fit templates spanning a few lines of code and are not designed for defects whose evidence is distributed across modules. As Guo et al.~\cite{guo2024outside} observe, many commercial tools cannot operate on code fragments at all: the whole project, or at minimum a complete source file, is required. VulnSage~\cite{vulnsage2025} quantifies the consequence directly, finding that zero-shot detection accuracy varies materially with analysis granularity across 593 CVEs in large C/C++ projects. We therefore treat function-level results as non-comparable rather than as a baseline.

\subsection{Model-specific observations}

Model families exhibit distinct behaviors across the Top100 dataset. For instance, \emph{Claude} achieved the highest accuracy of 94.06\%, while discovering 8/8 CVE-assigned vulnerabilities at the cost of \$1.45 per scan. \emph{DeepSeek-Reasoner} excelled on Hist20 with guided strategies with a fraction of the cost of \$0.08 per scan, but underperformed on Top100 and caught 3/8 new CVEs, indicating that CWE-specific directives can be catalytic for some models. \emph{Gemini} managed to balance cost and quality at \$1.01 per scan and an 84.2\% accuracy, while catching 4/8 new CVEs. Finally, \emph{GPT} exhibited premature termination and the highest query failure rates up to 68\%, resulting in lower coverage, still catching 4/8 CVEs at the cost of \$0.22 per scan.
The consensus analysis revealed that reliable vulnerability detection required multiple model configurations, mostly focusing on models that can serve vulnerability detection capabilities. At least 3 configuration scans are recommended to recover more than 95\% of the detected CVEs. From a cost-effectiveness perspective, at an average cost of \$5 per 10-configuration package scans, QRS provides a substantial cost reduction compared to traditional security audits while maintaining the capability to discover CVE-worthy vulnerabilities in production libraries.

\section{Discussion \& Limitations}
\label{sec:discussion}

QRS generated hypotheses and CodeQL queries across 115 CWE categories on the Top100, with the highest concentration in Injection and Code Execution (CWE-502, CWE-094, CWE-078: 1369 findings), Path Traversal (CWE-022, CWE-073, CWE-059: 670 findings), and DoS (CWE-1333, CWE-400: 578 findings), highly dependent on package profiles.

LLMs' non-determinism affects reproducibility, as repeated scans may yield different results. However, multiple configurations converge toward consistent detection. QRS also seems to predictively scale to large codebases without degradation, as \textbf{Q} is a code slicer that effectively generates targeted queries rather than ingesting entire raw code files, while keeping the costs and execution times manageable.

Query reuse and caching could substantially reduce pipeline costs, as CodeQL operations are the most expensive. This does not eliminate the \textbf{Q} agent as queries are synthesized per target, but recurring patterns may suggest a hybrid deployment of cached generic layer queries without any cost on top of the contextualized ones.

Extending QRS from Python to Java required a Java-specific knowledge schema, a Java package profiling tool, and a build-aware CodeQL database creation step. No further changes to the architecture or pipeline logic were needed. There is a current limitation regarding cross-language applications, but this can be addressed by generating multi-language CodeQL databases.

Our controlled CWE-Bench-Java experiment revealed that CWE-class preprompts are critical for guided discovery, hinting that an archive of CWE templates is ideal for OWASP Top 10 or case-specific security audits. An archive of CWE-022, 078, 079, and 094 is already available in the artifacts repository.

Scan duration and costs scale with codebase size in a predictable manner \autoref{tab:ablation-cost}, which is what makes routine deployment practical.

\section{Background and Related Work}
\label{sec:related}

To contextualize the QRS framework, we review prior work in static analysis, CodeQL-based program analysis, and recent advances in agentic AI for cybersecurity.

\noindent\textbf{Static Analysis \& Taint Tracking.}
\label{subsec:background_static_analysis}

SAST examines source code to detect potential vulnerabilities without execution. Taint analysis, a core SAST technique, tracks data flows from untrusted sources to sensitive sinks and has been widely used to identify issues such as injection vulnerabilities and insecure data
handling~\cite{schwartz2010all}. Despite its effectiveness, static taint analysis often struggles with scalability and precision when applied to large or dynamically typed codebases, frequently leading to false positives or negatives~\cite{salis2021pycg}.

\noindent\textbf{Pre-LLM pattern and rule inference.}
Deriving detection logic automatically rather than authoring it by hand predates language models. Yamaguchi et al.~\cite{yamaguchi2014cpg} introduced CPGs, unifying AST, CFG, and PDG information into a single queryable structure, and subsequently showed that search patterns for taint-style vulnerabilities can be \emph{inferred} from a codebase by clustering definition--use chains and expressed as graph traversals~\cite{yamaguchi2015}. A parallel line targets
\emph{recurring} vulnerabilities by matching against known vulnerable code rather than modeling vulnerability classes; VMud~\cite{vmud2024} selects critical fixing functions and performs semantically equivalent statement matching to detect clones of known defects.

These approaches share QRS's premise, that expert-authored rules are the bottleneck, and differ in where the missing knowledge comes from. Yamaguchi et al.\ mine it from statistical regularities within the target codebase, which requires the vulnerable pattern to recur often enough to form a cluster. VMud mines it from known patches, which requires the vulnerability to be known somewhere already. QRS elicits it from a model's pretrained semantic knowledge conditioned on package structure, removing both requirements at the cost of non-determinism and hallucination risk, which are the failure modes that \textbf{R} and \textbf{S} exist to contain.

\noindent\textbf{CodeQL \& Code Analysis.}
\label{subsec:background_codeql} CodeQL~\cite{CodeQL}, developed by GitHub to automate code security scanning, represents a state-of-the-art tool in semantic code analysis. It supports most major programming languages and can be configured to automatically scan all public GitHub repositories. Hence, it can significantly enhance the security of open-source software. Unlike rule-based pattern matchers, CodeQL converts code into abstract, high-level objects and control flows, stripping out non-semantics, and storing the results in a relational database. Researchers can query the database using QL~\cite{ql_language} in an SQL-like manner to detect sources, sinks, and flow configurations. While powerful, CodeQL has a steep learning curve, making it difficult to write effective queries without deep expertise in QL's syntax and the target library's API.

\noindent\textbf{Agentic AI.}
\label{subsec:background_agentic_ai}
LLMs have demonstrated remarkable capabilities in contextualizing code, yet suffer from `hallucinations', generating potentially plausible but incorrect facts. AI agents extend LLMs when equipped with \emph{tools} and \emph{reasoning iterations}. With tools, LLMs can execute functions that allow the model to interact with the environment or retrieve knowledge not directly available~\cite{claudetools}, e.g., retrieving online content, executing system commands, and ingesting their output. On the other hand, reasoning iterations can be considered chain-of-thought operations that enable running agents to plan, execute, and observe outputs to refine their next actions~\cite{claudereasoning,claudeagents}.

Nevertheless, one of the big advantages of LLMs is their unpredictability. Temperature configurations affect how deterministic or creative agents can be. For code generation or code review tasks, values between 0 and 1 are usually recommended~\cite{deepseek_temperature_settings,gemini_temperature_settings,claude_temperature_settings,gpt_temperature_settings}.

\subsection{LLMs in Vulnerability Analysis} 

The intersection of LLMs and static analysis for vulnerability detection has emerged as a rapidly evolving research area. The landscape can be divided into four categories, each with distinct trade-offs between automation, contextual reasoning, and practical applicability.

\noindent\textbf{Pattern-Based Classifiers.}
Discriminative approaches such as VulBERTa~\cite{hanif2022vulberta}, VulSim~\cite{vulsim2024}, and ReGVD~\cite{nguyen2022regvdrevisitinggraphneural} employ transformer architectures or contrastive learning to classify code. While achieving reasonable accuracy on curated benchmarks, these methods require extensive labeled training data and struggle with novel vulnerability patterns unseen during training, thereby limiting their effectiveness on real-world codebases. Finer-grained variants operate below the function level: LineVD~\cite{linevd2022} frames statement-level detection as a node classification task on a program dependence graph. Independently of granularity, this family shares an evaluation assumption QRS does not adopt, namely that a vulnerability is localizable within a pre-extracted code unit supplied by the benchmark. We therefore evaluate on complete packages with no pre-supplied localization, which is a strictly harder setting.

\noindent\textbf{LLM-Guided Static Analysis.}
Other hybrid approaches, such as LLMxCPG~\cite{lekssays2025llmxcpg} and GPTScan~\cite{sun2024gptscan}, combine CPGs with LLM reasoning to identify vulnerability-relevant code slices. However, these frameworks operate in a single-pass manner, without iterative refinement or recovery from initial misclassifications, slicing, or querying failures.

\noindent\textbf{Multi-Agent Agentic Frameworks.}
Recent work explores multi-agent architectures for vulnerability detection. LLM4Vuln~\cite{sun2025llm4vulnunifiedevaluationframework} demonstrates that external knowledge enhancement benefits logic-heavy vulnerabilities but may harm detection in traditional languages. VulAgent~\cite{wang2025vulagenthypothesisvalidationbasedmultiagent} introduces hypothesis validation with specialized agents for multi-view detection, achieving 26.6\% pairwise correct prediction on PrimeVul. Notably, they report reducing FP rates by 36\% through structured hypothesis pruning.

Concurrently with this work, SemTaint~\cite{semtaint2026} applies a three-agent design (source and sink identification, call-graph edge resolution, and flow-summary refinement) to extract package-specific taint specifications for JavaScript, feeding them to CodeQL and detecting 106 of 162 npm vulnerabilities that CodeQL alone could not, plus four novel findings. SemTaint is the closest architectural neighbor to QRS: both are multi-agent, both use CodeQL as the symbolic substrate, and both target the package ecosystem of a dynamically typed language. The distinction lies in what the agents produce. SemTaint's agents produce \emph{specifications} that parameterize CodeQL's existing query suite; QRS's \textbf{Q} agent produces the \emph{queries themselves}, including their dataflow configurations, from a syntax schema containing no vulnerability rules. The two are complementary: SemTaint's flow summaries would directly improve QRS's handling of taint crossing third-party boundaries, a known limitation of our Python evaluation.

IRIS~\cite{iris2025} is a taint specification engine that leverages LLMs for the specifications and augments CodeQL. However, IRIS requires explicit CWE specification as input and relies on CodeQL's call graph to connect sources and sinks. MoCQ~\cite{mocq2025} is a neuro-symbolic framework that generates vulnerability patterns via LLMs and translates them into detection queries. While MoCQ focuses on pattern generalization, QRS emphasizes per-package deep analysis with iterative hypothesis refinement, enabling the detection of package-specific vulnerabilities.

LARA~\cite{299627} is a static taint analysis framework for embedded systems that leverages semantic relations between frontend inputs (URIs and keys) and backend handling code, and combines pattern-based static analysis with LLM-aided semantic reasoning. In particular, LLMs are used to analyze decompiled backend code and infer the semantic roles of functions, such as identifying URI registration functions and key-handling routines, by capturing intent that is difficult to express through syntactic patterns alone. By intersecting LLM-derived results with rule-based analysis, LARA reduces false positives while retaining high coverage, enabling accurate identification of both hidden and non-hidden taint sources. Evaluated on real-world IoT firmware, LARA discovered 245 previously unknown vulnerabilities that were confirmed by vendors, with 162 assigned CVE identifiers.

For automated Query synthesis, QLCoder~\cite{qlcoder2025} is an agentic framework that generates CodeQL queries directly from CVE metadata, built upon Claude, and provides a custom MCP (Model Context Protocol) interface for syntactic guidance. The key distinction is that QLCoder requires CVE metadata inputs and focuses on reproducing known vulnerabilities. QLPro~\cite{qlpro2025} presents a framework for automated CodeQL rule generation for individual projects, employing LLMs with a triple-voting mechanism for classification. While QLPro excels at systematic API extraction and classification for Java projects, QRS takes a broader approach by generating complete logic, including inter-procedural data flows, without requiring API-level enumeration, and extends beyond Java. KNighter~\cite{knighter_yang_2025} synthesizes Clang Static Analyzers from Linux kernel patches, discovering 92 bugs (30 CVEs). The kernel code is targeted via Clang SA rather than CodeQL, but shares a similar philosophy of using LLMs for rule generation, rather than simply triaging alerts. KNighter learns from past patches whereas QRS self-generates hypotheses from a minimal syntax definition.

\noindent\textbf{Post-scan alert filtering.}
A body of work applies LLMs downstream of an existing analyzer to triage or suppress its alerts. LLM-SAST Genius~\cite{llmsastgenius2025} examines the synergy between traditional SAST and LLMs for alert triage. Klieber and Flynn~\cite{llmsalerts2025} study prompt design for adjudicating individual static-analysis alerts and report that models perform markedly better when asked about a specific issue type at a specific line than when asked to find issues freely, an observation that motivates our decision to hand \textbf{R} pre-localized candidates rather than raw files. LLM4FPM~\cite{llm4fpm2025} attacks the context problem directly, using an extended code property graph slicer and a file reference graph to supply line-precise, complete context for each warning. BugLens~\cite{buglens2025} guides an LLM through structured security-impact and constraint-assessment steps over taint-style warnings in the Linux kernel, improving precision roughly sevenfold and surfacing four previously unreported bugs.

All of these are bounded above by the recall of the analyzer they post-process: a defect the underlying tool never flags cannot be recovered. That is the constraint QRS removes by generating the detection logic itself. The comparison is one of position in the pipeline rather than superiority, since BugLens's precision on its target class exceeds ours. QRS's \textbf{S} agent is the closest analog to this line of work, and \autoref{tab:ablation-stages} isolates its contribution: applied to \textbf{R}'s output, it removes $55.7\%$ of false positives at a $4.7\%$ true-positive cost.

Finally, CVE-Genie~\cite{cvegenie2025} addresses the complementary problem of automatically reproducing known vulnerabilities through their CVE format to generate a high-quality vulnerability dataset using a multi-agent framework.

\section{Limitations}
\label{sec:threats-validity}

\noindent\textbf{What our metrics measure.} For Hist20 and CWE-Bench-Java, ground truth is external and detection rate carries its usual meaning. For Top100 it does not. No scanner can measure recall over a codebase whose full vulnerability set is unknown; the quantity we report
for Top100 is verdict retention (\autoref{tab:confusion-matrix}), the rate at which \textbf{S} preserves findings that human review confirmed. A vulnerability that \textbf{Q} never surfaced is invisible to this metric and is not counted as a false negative. Retention should therefore be read as a property of the expert review stages, not of the framework's coverage.

\noindent\textbf{Estimator choice.} Pooled (micro) and per-project (macro) F1 differ substantially when per-project positive counts are small: our Top100 corpus yields $\text{F1}_\mu = 0.503$ against $\text{F1}_M = 0.192$. We report both wherever we compare against prior work, and we compare only like estimators (\autoref{tab:framework-comparison}).

\noindent\textbf{TP/MR conflation.} Our custom Manual Review (MR) label denotes a True Positive (TP) of reduced confidence (50--89 on the confidence scale defined in the \textbf{R} prompt), not a distinct outcome class. We pool TP and MR when computing performance metrics.

\noindent\textbf{Ground-truth subjectivity.} Top100 labels derive from expert review by the authors, corroborated where possible by maintainer response (CVE assignment, patch, or documented acknowledgment). Maintainer response is a strong but not neutral signal: it reflects maintainer risk appetite as much as technical severity, as the acknowledged-but-not-assigned findings illustrate.

\noindent\textbf{Sanitize-stage framing bias.} \textbf{S} does not operate on an entirely clean slate. It receives \textbf{R}'s artifact, including \textbf{R}'s verdict, confidence and reasoning, and is therefore anchored by them. We describe \textbf{S} as an independent \emph{judge} of \textbf{R}'s evidence, not an independent \emph{detector}. A blinded \textbf{S} variant that receives evidence without \textbf{R}'s verdict would isolate this effect and is left to future work.

\noindent\textbf{Duplicate accounting.} The same underlying defect can surface through several \textbf{Q} queries and appear in several MITRE clusters. We deduplicate statically before handoff to \textbf{S} and before all reported metrics. Defects manifesting at different lines of the same flow may still be double-counted.

\noindent\textbf{Non-determinism.} Repeated scans with identical configurations do not produce identical outputs. We mitigate this by reporting across 10 configurations and 1{,}000 scans rather than single runs, but individual detection claims should be read as existence results, not as reproducible expectations.

\noindent\textbf{Sample selection.} Top100 selects for popularity and active maintenance, which biases \emph{against} finding vulnerabilities: these are the most-scrutinized packages in the ecosystem. Detection rates on unmaintained or less-scrutinized code are likely higher and are not measured here. Conversely, the packages are unrepresentative of proprietary enterprise codebases in size, architecture, and dependency structure.

\noindent\textbf{Benchmark contamination.} Hist20 consists of published CVEs that predate every evaluated model's training cutoff and must be treated as a functional sanity check, not as evidence of discovery capability; we de-emphasize its metrics accordingly. The Top100 result is
not subject to this concern for the 8 CVE-assigned and 4 acknowledged findings, which were undisclosed at scan time and therefore absent from any training corpus. The rediscovered CVEs can be considered contaminated under the same principle; we mitigate by disabling the \texttt{search\_cve} tool until after \textbf{R} completes, and note that detection proceeds through CodeQL queries evaluated against a locally built AST/dataflow database, not through recall of advisory text. We cannot rule out the possibility that a model's prior exposure to a package's patch history steers hypothesis generation, yet this does not prove that pre-trained data can steer models toward proper queries.

\noindent\textbf{Rule-free, not schema-free.} QRS operates without predefined \emph{vulnerability} rules, but not without structure. It depends on the CodeQL QL schema, the standard library's \texttt{RemoteFlowSource}, \texttt{CodeExecution} and similar concept classes, and on the canonical patterns in \texttt{knowledge.md}. These encode substantial prior human knowledge about what a vulnerability looks like. Our claim is the elimination of per-vulnerability rule authorship, not the elimination of the analysis framework.

\subsection{Reproducibility}

Model APIs are versioned, deprecated or silently updated. We record exact model identifiers and execution timestamps (\autoref{tab:model-configuration-matrix}) and release all prompts, knowledge schemas, generated queries and outputs, but exact numerical reproduction against future model versions should not be expected. Cost figures assume no provider-side prompt caching and therefore represent an upper bound; observed costs with caching enabled will be lower by an amount we cannot measure from the client side.

\section{Conclusion \& Future Work}
\label{sec:conclusion}
This work introduces QRS, a neuro-symbolic framework that inverts the dominant paradigm for LLM-assisted vulnerability detection. Most prior work applies language models as post hoc filters to rule-based scanner outputs and fine-tuned models. QRS employs LLMs as the generative core to synthesize CodeQL queries from a minimal schema, perform semantic validation through iterative code inspection, and produce PoC exploits, without predefined rules, package knowledge, fine-tuning, or training data beyond a few-shot examples.

Our evaluation supports multiple crucial findings. First, QRS achieves high accuracy and recall on vulnerabilities: a combined configuration set detected 19/20 historical CVEs in our Hist20 dataset, and reached 90.62\% accuracy, 86.96\% precision, and 100.00\% recall. Second, QRS discovers novel and complex vulnerabilities in production code that existing tooling misses. Scanning the 100 most-downloaded PyPI packages, which are already being monitored by GitHub's Advanced Security with CodeQL, Dependabot, and AI Copilot, QRS identified 41 vulnerabilities, 8 of which were previously unknown that received 8 CVEs, and 4 maintainer-acknowledged issues. Third, QRS substantially reduces alert fatigue. Our noisiest configuration achieved a 64.8\% reduction and 93.49\% classification accuracy. Opengrep and Bandit on the same corpus produced more than an order of magnitude more alerts, whereas CodeQL yielded no relevant findings. The above is also validated in Java, demonstrating that QRS can be easily adapted to another language. In fact, QRS outperforms IRIS in both statistics and absolute numbers, even in a reduced dataset. The framework operates at a low cost across model families, making comparisons favorable to manual security audits while preserving the interpretability that automated approaches typically sacrifice. Moreover, every finding includes query provenance, data-flow evidence, and PoC exploit code. Our experiments also showed that model-specific behaviors affect coverage. Finally, our ground truth for the Top100 dataset relies on expert review and maintainers' feedback.

The architectural separation of QRS agents extends to cross-language analysis. CodeQL's unified representations for supported languages require a schema adaptation, not a fundamental redesign, and further extensions will be implemented in the future. More sophisticated clustering that spans CWE boundaries may reveal multi-stage vulnerabilities that are invisible to taxonomy-driven grouping. We believe that replacing the sequential pipeline with continuous inter-agent feedback via directed graph, or swarm approaches, could reduce verdict inconsistencies; therefore, we will explore this in future work. Also, hybrid query caching and common anti-patterns is a gap for potential future work. Finally, we plan to integrate agentic dynamic validation by executing generated exploits in sandboxed environments to provide stronger evidence of exploitability than static reasoning alone can.

\section*{Acknowledgment}
This work was supported by the European Commission under the Horizon Europe Programme, as part of the projects SafeHorizon (Grant Agreement No. 101168562) and CYMEDSEC (Grant Agreement No. 101094218). 

The content of this article does not reflect the official opinion of the European Union. Responsibility for the information and views expressed therein lies entirely with the authors.

\appendix

\section{Failed CVE-2021-32677 detection}
CVE-2021-32677 (FastAPI open redirect) was not detected by any model configuration. Manual inspection revealed that the vulnerability requires reasoning about HTTP redirect semantics and framework-specific routing behavior, patterns not captured by our current QL knowledge base. More precisely, the vulnerability stems from the fact that requests with content type text/plain are exempt from CORS preflights. The latter does not match any specific code pattern. However, extending the knowledge schema of QRS with web framework routing primitives could potentially identify further vulnerabilities.
 
\section{Configurations}

\begin{table}[ht]
\centering
\caption{Model configuration matrix}
\label{tab:model-configuration-matrix}
\footnotesize
\rowcolors{2}{gray!10}{white}
\begin{tabular}{@{}lllcc@{}}
\toprule
\textbf{Short} & \textbf{Q Model} & \textbf{R + S Models} & \textbf{Temp} & \textbf{Iterations} \\
\midrule
C$_0$ & claude-sonnet-4-5 & claude-haiku-4-5 & 0.0 & 20+10 \\
C$_1$ &claude-sonnet-4-5 & claude-haiku-4-5 & 1.0 & 20+10 \\
DC$_0$ &deepseek-chat & deepseek-chat & 0.0 & 20+10 \\
DC$_1$ &deepseek-chat & deepseek-chat & 1.0 & 20+10 \\
DR$_0$ &deepseek-reasoner & deepseek-reasoner & 0.0 & 20+10 \\
DR$_1$ &deepseek-reasoner & deepseek-reasoner & 1.0 & 20+10 \\
G$_0$ & gemini-3-pro-preview & gemini-2.5-pro & 0.0 & 20+10 \\
G$_1$ & gemini-3-pro-preview & gemini-2.5-pro & 1.0 & 20+10 \\
GPT$_0$ &gpt-5.1 & gpt-4.1 & 0.0 & 20+10 \\
GPT$_1$ &gpt-5.1 & gpt-4.1 & 1.0 & 20+10 \\
C$_3$ &claude-sonnet4-6 & claude-sonnet-4-6 & 1.0 & 20+20 \\
\bottomrule
\end{tabular}
\end{table}

\section{Prompts \& Logs}
We release anonymized versions of system prompts, preprompts, and knowledge schemas used in the evaluation.

\begin{lstlisting}[
  caption={\textbf{Q agent}: Recovery subroutine},
  label={lst:q-agent-iteration},
  style=geodataviz,language=Markdown
]
[DeepSeek]: Excellent! I found 4 instances of `__import__` being used in the constructor.py files. This is a critical finding as it indicates the library can dynamically import modules during YAML parsing. Let me now search for the specific YAML tag handlers that enable arbitrary Python object construction and for YAML tag registration patterns that allow object instantiation.
INFO - [*] Processing tool call: execute_codeql_query
INFO - [*] Executing CodeQL query with
[*] Severity: critical
[*] CWE: ['cwe-502', 'cwe-094']
[*] Hypothesis: Find YAML tag handlers that construct Python objects, particularly those handling 'python/object' or 'python/object/apply' tags which can instantiate arbitrary classes
INFO - [*] Running command: codeql database analyze 
databases/pyyaml-5.3-db 20260118133453.ql \
  --format sarif-latest \
  --output reports/pyyaml-5.3/20260118133453_results.sarif
ERROR - [-] CodeQL query execution failed: CodeQL query compilation failed - no output generated
[DeepSeek]: Let me fix that query.
INFO - [*] Processing tool call: fixup_codeql_query
INFO - [*] Attempting to fix CodeQL query.
INFO - [*] Running command: codeql query compile 709dfdb5-d62d-4cfa-8ac7-bcd3d85835f9.ql
INFO - [+] Query is valid (attempt 1/3).
INFO - [+] CodeQL query fixed successfully.
[DeepSeek]: Good, the query is valid. Let me execute it.
\end{lstlisting}

\begin{lstlisting}[
  language={json},
  caption={\textbf{R agent}: Artifact with evidence},
  label={lst:r-agent-finding},
  style=geodataviz
]
{
 "codeql_query_uuids": ["20260118133427", "20260118133627"],
 "finding_id": "21864d8a-a727-41b2-98ef-a42caad42b3b",
 "verdict": "tp",
 "confidence": 95,
 "severity": "critical",
 "cwe_ids": ["cwe-502", "cwe-094"],
 "description": "Unsafe __import__ in find_python_module allows arbitrary module import during YAML parsing with UnsafeConstructor [...]",
 "file_path": "lib/yaml/constructor.py",
 "start_line": 520,
 "start_column": null,
 "code_snippet": "def find_python_module(self, name, mark, unsafe=False)[...]",
 "reasoning": "The find_python_module method uses __import__(name) when unsafe=True. This is called by UnsafeConstructor.find_python_module which overrides the parent method with unsafe=True. The UnsafeConstructor is used by the default Loader class, making this ulnerability accessible through the standard yaml.load() API. [...]",
 "evidence": {
    "code_context": "The find_python_module method at line 520 uses __import__(name) when unsafe=True. UnsafeConstructor overrides this at line 716-717 to call the parent with unsafe=True. [...]",
    "dataflow_analysis": "YAML input -> tag:yaml.org,2002:python/module: -> construct_python_module -> find_python_module(unsafe=True) -> __import__(name)",
    "notes": "The vulnerability is in the default Loader class which uses UnsafeConstructor. Users who call yaml.load() without specifying a Loader will use this unsafe constructor by default."
  },
 "exploit_payload": {
    "code_snippet": "[...]",
    "scenario": "Attacker provides malicious YAML file that uses python/module tag to import arbitrary modules [...]",
    "likelihood": "high",
    "impact": "high",
    "exploitability": "easy",
    "prerequisites": "Application must use yaml.load() or yaml.Loader with untrusted YAML input",
    "category": "Remote code execution via unsafe deserialization"
  },
 "group": "critical_often_medium"
}
\end{lstlisting}

\begin{lstlisting}[
  language=Python,
  caption={\textbf{R agent}: YAML deserialization PoC exploit},
  label={lst:yaml-exploit},
  basicstyle=\ttfamily\scriptsize,
  style=geodataviz
]
import yaml

malicious_yaml = "!!python/module:os"
result = yaml.load(malicious_yaml, Loader=yaml.Loader)
\end{lstlisting}

\begin{lstlisting}[
  language=sql,
  caption={\textbf{Q agent}: Generated query to detect dangerous code execution},
  label={lst:dangerous-exec},
  style=geodataviz
]
/**
 * @name Find dangerous code execution methods
 * @kind problem
 * @problem.severity warning
 * @id py/find-dangerous-code-execution
 */
import python

predicate isDangerousFunction(string name) {
  name in ["eval", "exec", "compile", "__import__"]
}

from Call c, string funcName
where (c.getFunc().(Name).getId() = funcName
    and isDangerousFunction(funcName))
  or  (c.getFunc().(Attribute).getName() = funcName
    and isDangerousFunction(funcName))
select c, "Dangerous code execution function found: " + funcName
\end{lstlisting}

\begin{lstlisting}[
  caption={\textbf{R agent}: Extra Strategy Definitions},
  label={lst:r-review-priorities},
  style=geodataviz,language=Markdown
]
### Review Priority
When reviewing CodeQL findings prioritize validation of:

#### Code Execution & Deserialization Vulnerabilities
- **CWE-94**: Code Injection vulnerabilities
- **CWE-502**: Unsafe deserialization patterns
- **CWE-20**: Input validation issues leading to code execution

### Key Validation Criteria
#### 1. Exploitability Assessment
For each finding, thoroughly investigate:
**Can attacker control the input?** Is the YAML input from untrusted sources?
**What can be achieved?** Can arbitrary code execution be triggered through the finding?
**What are the prerequisites?** Which loading functions enable the vulnerability?
**Is it in the critical path?** Does it affect common/default usage patterns?

#### 2. Context Analysis
Carefully examine:
**Code flows**: Trace from YAML input -> parsing -> object construction -> execution
**Function semantics**: Understand what each loading function does and its safety model
**Default behavior**: Is the vulnerability present in default/recommended usage?
**Attack vectors**: How would a real attacker craft malicious YAML to exploit this?

#### 3. True Positive Indicators
Strong signals for TRUE POSITIVE classification:
Direct path from user-controlled YAML to code execution primitives
Functions that deserialize and instantiate Python objects from YAML
Dynamic code execution (`__import__`, `eval`, `exec`, object instantiation) triggered by parsed data
Minimal prerequisites for exploitation
Affects documented/common API usage patterns

#### 4. False Positive Indicators
Be cautious about marking as FALSE POSITIVE if:
The code is in a constructor/loader component
Default behavior enables the vulnerability
Common usage patterns are affected
The library is designed to parse untrusted data
\end{lstlisting}

\begin{lstlisting}[
  language=json,
  caption={\textbf{Q agent}: Package profile for PyYAML},
  label={lst:package_profile},
  style=geodataviz
]
{
  "metadata": {
    "name": "PyYAML",
    "version": "5.3",
    "description": "YAML parser and emitter for Python",
    "long_description": "YAML is a data serialization format designed for human readability\nand interaction with scripting languages[...]",
    "license": "MIT",
    "platforms": "Any",
    "classifiers": [
      "Programming Language :: Python :: 3.8",
      "Programming Language :: Python :: Implementation :: CPython",
      "Programming Language :: Python :: Implementation :: PyPy",
      "Topic :: Software Development :: Libraries :: Python Modules",
      "Topic :: Text Processing :: Markup", [...]
    ],
    "package_dir": {},
    "packages": ["yaml"],
    "ext_modules": [null],
    "python_requires": ">=2.7, !=3.0.*, !=3.1.*, !=3.2.*, !=3.3.*, !=3.4.*"
  },
  "metadata_source": "setup.py",
  "files": [
    {
      "path": "lib3/yaml/constructor.py",
      "lines": 610
    },
    {
      "path": "lib3/yaml/loader.py",
      "lines": 47
    }, [...]
  ],
  "total_files": 35,
  "total_lines": 9269,
  "imports": [
    "os", "parser",
    "re", "reader",
    "serializer", [...]
  ]
}
\end{lstlisting}

\begin{lstlisting}[
  caption={CodeQL knowledge base},
  label={lst:codeql-knowledge},
  style=geodataviz,language=Markdown
]
## Core Requirements
- Running CodeQL CLI v2.23.2 or later
- **ALWAYS** include @kind metadata property
- For @kind path-problem: MUST include @problem.severity AND import PathGraph
- For @kind problem: MUST include @problem.severity (values: error, warning, recommendation, note)
- Do NOT use @severity metadata property - it's deprecated and causes warnings
- Path-problem select format: `select sink.getNode(), source, sink, "Message $@.", source, "label"`
[...]
## Data Flow Nodes
- Use `DataFlow::CallCfgNode` for function/method calls (NOT deprecated DataFlow::CallNode)
- Use `DataFlow::ExprNode` for expressions: `DataFlow::exprNode(expr)`
- Use `DataFlow::ParameterNode` for function parameters
- For local sources: Use `DataFlow::LocalSourceNode` and `.flowsTo()` predicate
- Convert between: `node.asExpr()` and `node.asCfgNode()`

## Sources and Sinks
- Remote sources: `source instanceof RemoteFlowSource` (from RemoteFlowSources module)
- File access sinks: `sink = any(FileSystemAccess fa).getAPathArgument()`
- Command execution: `sink = any(SystemCommandExecution cmd).getACommandArgument()`
- SQL execution: `sink = any(SqlExecution sql).getAnArgument()`
- Code execution: `sink instanceof CodeExecution`
- Path argument access: `call.getArg(0)` or `call.getParameter(0, "param_name")`

## Local Data Flow
- Local flow (within function): `DataFlow::localFlow(source, sink)`
- Local taint (with non-value-preserving steps): `TaintTracking::localTaint(source, sink)`
- Get local sources of node: `node.getALocalSource()`
- LocalSourceNode flow: `sourceNode.(DataFlow::LocalSourceNode).flowsTo(sinkNode)`
\end{lstlisting}
\begin{lstlisting}[
  caption={\textbf{S agent}: Finalized verdict},
  language=json,
  label={lst:s-artifact-output},
  style=geodataviz
]
"sanitization_check": {
  "verdict": "tp",
  "confidence": 92,
  "recommendation": "accept",
  "concerns": false,
  "flags": [
    {
      "type": "ZERO_DAY",
      "severity": "critical",
      "message": "High-confidence exploitable vulnerability in production serialization library",
      "evidence": "UnsafeConstructor.add_multi_constructor registers python/object/apply tag which enables arbitrary code execution through object instantiation. Constructor is aliased to UnsafeConstructor, making this available in default Loader. The construct_python_object_apply method calls make_python_instance with unsafe=True, allowing instantiation of arbitrary classes with attacker-controlled arguments."
    }
  ],
  "reasoning": "This is a legitimate and critical vulnerability in PyYAML's default Loader. The code trace is accurate: the python/object/apply tag handler is registered on UnsafeConstructor, which is aliased as Constructor and used by the default Loader, enabling RCE through arbitrary object instantiation with attacker-controlled arguments[...]"
}
\end{lstlisting}

\begin{lstlisting}[
  caption={\textbf{Q agent}: Extra Strategy Definitions},
  label={lst:q-priority-areas},
  style=geodataviz,language=Markdown
]
### Priority Areas
Focus your security analysis on the following high-priority vulnerability classes:

#### 1. Code Execution Vulnerabilities
- **CWE-94**: Improper Control of Generation of Code ('Code Injection')
- **CWE-502**: Deserialization of Untrusted Data
- Look for patterns where code or objects can be instantiated from 
serialized/parsed data

#### 2. Unsafe Deserialization Patterns
- **CWE-20**: Improper Input Validation
- Examine how the library handles YAML tags and type annotations
- Investigate object construction, module imports, and dynamic code execution during parsing

#### 3. Dynamic Code Execution
- Functions that evaluate, execute, or import code dynamically (`eval`, `exec`, `__import__`, `compile`)
- Object instantiation from strings or parsed structures
- Python-specific YAML tags that enable object creation
\end{lstlisting}

\lstset{
  basicstyle=\ttfamily\footnotesize,
  breaklines=true,
  columns=fullflexible
}

\begin{table}[htbp]
\end{table}
\begin{lstlisting}[
  caption={\textbf{Q} System Prompt},
  label={tab:q-agent-prompt},
  style=geodataviz,language=Markdown
]
## Role
You are an autonomous security analysis research agent, specialized in analyzing PyPI packages for vulnerabilities.

## Goal
Discover security vulnerabilities in the target package via CodeQL and other relevant tools.

## Tools
- get_package_profile_info: Retrieve a profile of the target package. Prefer to start your analysis with this tool as it will include important metadata about the package structure, dependencies and files and run only once in the beginning.
- execute_codeql_query: Execute CodeQL queries against the package database to find vulnerabilities. This is your PRIMARY tool. If the generated query returns too many results, it may be too broad or noisy so refine it based on previous findings.
- fixup_codeql_query: Fix CodeQL queries that fail validation without consuming your iteration budget. Use this to correct errors.
- finalize_report: Finalize and submit your security analysis report when analysis is complete.
**YOU SHOULD:**
- Call `fixup_codeql_query` with the failing query code and error message

## Strategy
1. Start by retrieving the package profile to understand its structure and dependencies.
2. Prioritize critical, high, medium, low severity vulnerabilities in that order.
3. Formulate hypotheses about vulnerabilities. You can be broader initially, and then narrow down based on iteration budget.
4. Assign CWE IDs (one or more relevant) to your hypotheses when possible.
5. Write precise CodeQL queries to test your hypotheses.
6. **CRITICAL** If you come across CodeQL query validation issues prefer to run the fixup_codeql_query tool as it does not consume your iteration budget. Avoid wasting iterations on invalid queries and try to fix them for each of your hypotheses.
7. When confident you have covered major vulnerability classes, finalize your report.

### CodeQL Query Examples
#### Path-problem Query:
```ql
/**
 * @name Code injection via eval/exec
 * @kind path-problem
 * @problem.severity error
 * @id py/code-injection
 * @tags security external/cwe/cwe-094
 */

import python
import semmle.python.dataflow.new.DataFlow
import semmle.python.dataflow.new.TaintTracking
import semmle.python.dataflow.new.RemoteFlowSources
import semmle.python.Concepts

module CodeInjectionConfig implements DataFlow::ConfigSig {{
  predicate isSource(DataFlow::Node source) {{
    source instanceof RemoteFlowSource
  }}

  predicate isSink(DataFlow::Node sink) {{
    sink = any(CodeExecution ce).getCode()
  }}
}}

module CodeInjectionFlow = TaintTracking::Global<CodeInjectionConfig>;
import CodeInjectionFlow::PathGraph

from CodeInjectionFlow::PathNode source, CodeInjectionFlow::PathNode sink
where CodeInjectionFlow::flowPath(source, sink)
select sink.getNode(), source, sink, "Code injection from $@.", source.getNode(), "user input"
```

#### Problem Query:
```ql
/**
 * @name Unsafe deserialization
 * @kind problem
 * @problem.severity error
 * @id py/unsafe-deserialization
 * @tags security external/cwe/cwe-502
 */

import python
import semmle.python.ApiGraphs

from DataFlow::CallCfgNode call
where
  (
    call = API::moduleImport("pickle").getMember(["load", "loads", "Unpickler"]).getACall() or
    call = API::moduleImport("yaml").getMember(["load", "unsafe_load"]).getACall() or
    call = API::moduleImport("marshal").getMember(["load", "loads"]).getACall() or
    call = API::moduleImport("shelve").getMember("open").getACall()
  )
select call, "Potentially unsafe deserialization operation that could execute arbitrary code."
```

## Budget
Iteration budget: {max_iterations} iterations total. When you reach this limit, you MUST finalize your report.
**IMPORTANT**: Try using as much of your iteration budget as possible to ensure a thorough analysis.

## Notes
- You have full autonomy to decide your analysis strategy. Work efficiently and thoroughly within the iteration budget.
- You can revisit and refine or retry previous hypotheses if you are within iteration budget or you think your query was incorrect, inefficient or ineffective based on its results.
- If you think you are on the right track, continue refining your hypotheses and queries to pinpoint a potential finding.
- At each iteration, you will be informed of your current progress and remaining iterations.
- When you have only a few iterations remaining (1-2), prioritize finalizing your findings over new hypotheses.
\end{lstlisting}

\begin{lstlisting}[
  caption={\textbf{R} Agent system prompt configuration for vulnerability validation and exploit generation},
  label={tab:r-agent-prompt},
  style=geodataviz,language=Markdown
]
## Role
You are an expert autonomous security findings reviewer.

## Goal
Review security findings and submit INDIVIDUAL confirmed vulnerabilities with precise location details.

## Tools
- extract_code_snippet: Get code context for specific locations
- grep_search: Grep search for specific patterns in files
- trace_flow: Trace data flow from source to sink (use when codeFlows available). If you already have codeFlows, no need to call this tool. If you decide to use this tool prefer the default line_tolerance value as higher values tend to generate noise. If too many values are fetched try reducing the line_tolerance.
- submit_finding: Submit EACH individual confirmed vulnerability with precise location
- complete_review: Signal completion when you're done reviewing all findings in this group

## Instructions
- You are given a GROUP of related CodeQL query results.
- Each query may have MULTIPLE individual results at different file locations.
- Your job is to submit EACH CONFIRMED VULNERABILITY SEPARATELY with specific file/line information.
- DO NOT submit one verdict per query - submit one finding per SPECIFIC VULNERABLE LOCATION.
- The findings include independent CodeQL query results still relevant to each other via group context clustering (severity, prevalence, likelihood, CWEs) to help you catch even more complex vulnerabilities.
- Leverage the MITRE values for insights across relevant results.
- If you discover NEW vulnerabilities during your review, submit them too with full details.
- Use available tools to investigate raw code, data flows, potential sanitizers and/or mitigators.

## Review Process

1. Examine each query's results array - each result is a potential vulnerability at a specific location
2. Investigate suspicious locations using extract_code_snippet and trace_dataflow_path
3. For EACH confirmed vulnerability, call submit_finding with:
   - codeql_query_uuids: Array of query UUIDs this finding relates to (from the query results you examined). Use [] for new findings you discovered independently
   - Exact file path (from result.locations[0].physicalLocation.artifactLocation.uri)
   - Exact line numbers (from result.locations[0].physicalLocation.region)
   - Specific reasoning for THIS location
   - Evidence object containing your investigation results (code snippets, dataflow analysis, relevant context)
   - Dataflow_path array: If you called trace_dataflow_path, copy the 'results' array from the tool response directly into this field (do NOT summarize, include the raw results)
   - exploit_payload object: REQUIRED for TRUE POSITIVES (tp verdict) - generate working exploit with all required fields based on actual code context

## Important Notes
- Submit findings for true positives (tp), false positives (fp), or items needing manual review (mr)
- Focus on confirming the most critical/dangerous instances first
- You can discover NEW vulnerabilities during investigation and submit them too
- Use MITRE context (severity, prevalence, likelihood, related CWEs) to guide your analysis
- ALWAYS populate codeql_query_uuids with the UUIDs of queries this finding relates to
- ALWAYS try to populate as much detail as possible in your findings submissions into their respective fields (eg. severity, verdict, confidence, cwe_ids, description, file_path, start_line, code_snippet, reasoning)
- ALWAYS populate the file_path, start_line and code snippet with the exact vulnerable code
- ALWAYS populate the evidence field with investigation results - include code snippets you extracted, dataflow analysis, and any relevant context that supports your verdict
- CRITICAL: If you call trace_dataflow_path, you MUST copy the complete 'results' array from the tool response into the dataflow_path field - do not write a summary, copy the actual results
- CRITICAL: For TRUE POSITIVE findings (tp verdict), you MUST include exploit_payload with ALL required fields:
  * code_snippet: Working exploit code based on the ACTUAL vulnerable code you investigated (not generic PoC)
  * scenario: Short sentence describing the attack scenario (e.g., "Attacker uploads malicious YAML file to trigger code execution")
  * likelihood: low/medium/high based on actual code context and accessibility
  * impact: low/medium/high based on what the exploit achieves
  * exploitability: easy/medium/hard based on complexity of exploitation
  * prerequisites: What an attacker needs (e.g., "Authentication required" or null if none)
  * category: What this exploit achieves (e.g., "Remote code execution", "SQL injection", "Path traversal")
- Your exploit code should leverage the actual code patterns and vulnerabilities you discovered during investigation
- Base exploit_payload on the code context you have already loaded - you don't need to make additional calls
- More detailed fields = better quality findings
- If you spot build or test files that are not actually exploitable classify them as false positives (fp)

## Strategy
- You have full autonomy to decide your review strategy. Work efficiently and thoroughly within the iteration budget. Your iteration budget is {max_iterations} iterations per group of findings.
- When iterations are low (1-2 remaining), wrap up soon and submit remaining findings.
- On your FINAL iteration, you MUST submit findings for any remaining confirmed vulnerabilities.
- When you have completed your review, you can call complete_review to avoid consuming tokens.
\end{lstlisting}

\begin{lstlisting}[
  caption={\textbf{S} Agent system prompt configuration},
  label={lst:s-agent-prompt},
  style=geodataviz,language=Markdown
]
# Role
You are a Static Analysis Security expert validator.

# Goal
You receive a single finding and you must analyze it for accuracy and relevance.

# Important Guidelines
Your response MUST be a valid JSON object with the exact structure specified in the user prompt. Do not include any text before or after the JSON object as this will fail parsing.

## Tools
- verify_finding: Verify the verdict and confidence score and adjust according to Verdict, Confidence and Recommendation Guidelines
- flag_finding: Add the corresponding flags to each verified finding according to Flag Guidelines

## Strategy
**Validation Tasks:**
1. **Code Trace Accuracy**: Does the code snippet actually contain the claimed vulnerability? Look for hallucinations.

2. **Code Context Classification**: Analyze the file path and code to determine:
- Is this BUILD_SCRIPT code (setup.py, build configuration, installation scripts)?
- Is this TEST_CODE (test files, test utilities, smoke tests, examples)?
- Is this production/runtime code?
- Is this an actual Zero Day vulnerability?

3. **True Exploitability**: Is this truly exploitable in the package's context? Consider:
- For libraries: Is this a caller responsibility or actual library vulnerability?
- For infrastructure: Is this reachable by external input?
- Build/test code is not exploitable in production

4. **Vulnerability vs Code Smell**: Is this a real security vulnerability or just a best practice issue?

**Response Format (JSON only):**
```
{{
"verdict": "tp|fp|mr",
"confidence": 0-100,
"recommendation": "reject|review|accept",
"concerns": true/false,
"flags": [
{{"type": "HALLUCINATION|BUILD_SCRIPT|TEST_CODE|EXPLOITABILITY|
CODE_SMELL|CONTEXT_MISMATCH|KNOWN_CVE|NOVEL_FINDING|
ZERO_DAY",
  "severity": "critical|warning|info",
  "message": "Brief explanation",
  "evidence": "Supporting details from code/context|Comma-delimited CVE Id(s)"}}
],
"reasoning": "Overall assessment in 2-3 sentences"
}}
```
**Verdict Guidelines**:
- tp: True positive - legitimate security vulnerability
- fp: False positive - not a real vulnerability (hallucination, code smell, test code, etc.)
- mr: Manual review - uncertain, needs human expert review

**Confidence Guidelines** (0-100):
- 90-100: Very high confidence in verdict
- 70-89: High confidence, recommend manual review
- 50-69: Moderate confidence, recommend manual review if it you are missing context and have indicators of a tp otherwise reject as fp 
- 0-49: Low confidence, reject as fp

**Recommendation Guidelines**:
- reject: High-confidence false positive (HALLUCINATION, BUILD_SCRIPT, TEST_CODE with critical flags, obvious CODE_SMELL)
- review: Needs human attention (EXPLOITABILITY concerns, CONTEXT_MISMATCH, moderate confidence, mr verdict)
- accept: High-confidence true positive or only informational flags (KNOWN_CVE, NOVEL_FINDING, ZERO_DAY)

**Flag Guidelines**:
- Use KNOWN_CVE flag if matching CVEs exist (potential duplicate, informational)
- Use NOVEL_FINDING flag if no CVE match found but legitimacy uncertain
- Use ZERO_DAY flag if this is a HIGH-CONFIDENCE novel, exploitable vulnerability not matching any CVEs
- Use HALLUCINATION flag if claimed code doesn't exist in snippet
- Use BUILD_SCRIPT flag for any build/installation/packaging code
- Use TEST_CODE flag for any test/example/smoke test code
- Use EXPLOITABILITY flag if vulnerability is not truly exploitable in context
- Use CODE_SMELL flag for best practice issues, not real vulnerabilities
- Use CONTEXT_MISMATCH flag if finding doesn't align with package type/layer
\end{lstlisting}

\begin{table}[htbp]
\footnotesize
\begin{tabularx}{\columnwidth}{@{}X@{}}
\toprule
\textbf{Log excerpt: Q Agent} \\ \midrule

\texttt{[*] Agentic Security Analysis Session: 1} \\
\texttt{[*] Target Package: pyyaml-5.3} \\
\texttt{[*] LLM Model: deepseek/deepseek-chat} \\
\texttt{[*] Temperature: 0.0, Max Tokens: 8000, Max Iterations: 20} \\
\texttt{[*] Loaded system prompt knowledge from knowledge.md} \\
\texttt{[*] Iteration 1/20} \\[1ex]

\textbf{[DeepSeek]:} I'll begin by retrieving the package profile to understand the structure of pyyaml-5.3 \\[1ex]

\texttt{[*] Processing tool call: get\_package\_profile\_info} \\
\texttt{[*] Retrieving package profile info...} \\
\texttt{[*] Profiling the package pyyaml-5.3.} \\
\texttt{[*] Running command: python3 profiler\_minimal.py} \\
\texttt{~~~~--path packages/pyyaml-5.3} \\
\texttt{~~~~--output reports/pyyaml-5.3/package\_profile.json} \\
\texttt{[*] Iteration 2/20} \\[1ex]

\textbf{[DeepSeek]:} Now I have a good understanding of the package. PyYAML is a YAML parser and emitter for Python. I should focus on:
\begin{itemize}[nosep, leftmargin=12pt, itemsep=1pt]
    \item YAML parsers can execute arbitrary code through Python-specific tags
    \item Through dynamic object creation or evaluation
    \item Improper handling of malicious YAML content
\end{itemize}
\vspace{1ex}

\texttt{[*] Processing tool call: execute\_query} \\
\texttt{[*] Executing CodeQL query with Severity: critical,} \\
\texttt{~~~~CWE: ['cwe-094', 'cwe-502'], Hypothesis: PyYAML may} \\
\texttt{~~~~contain dangerous function calls like eval, exec,} \\
\texttt{~~~~compile, \_\_import\_\_, or getattr...} \\
\texttt{[+] CodeQL query executed successfully.} \\
\texttt{[+] CodeQL query found 13 findings.} \\
\texttt{[*] Iteration 11/20} \\[1ex]

\textbf{[DeepSeek]:} Excellent! I found 13 dangerous function calls in the YAML module... \\[1ex]

\texttt{\{} \\
\texttt{~~"success": true,} \\
\texttt{~~"results": [...],} \\
\texttt{~~"codeFlows": [\{} \\
\texttt{~~~~"message": \{"text": "User input flows to} \\
\texttt{~~~~~~dangerous getattr in YAML constructor."\},} \\
\texttt{~~~~"threadFlows": [\{"locations": [\{} \\
\texttt{~~~~~~"physicalLocation": \{} \\
\texttt{~~~~~~~~"artifactLocation": \{} \\
\texttt{~~~~~~~~~~"uri": "lib3/yaml/constructor.py"\}\}\}]\}]\}]} \\[1ex]

\textbf{Current Status:} Iteration 14/20 (Remaining: 6). Queries executed: 9. Found 6 critical data flow paths to dangerous \texttt{\_\_import\_\_} and \texttt{getattr} calls. \\
\bottomrule
\end{tabularx}
\caption{Q Agent execution log excerpt for pyyaml-5.3 vulnerability discovery}
\label{tab:q-agent-log}
\end{table}

\begin{table}[htbp]
\footnotesize
\begin{tabularx}{\columnwidth}{@{}X@{}}
\toprule
\textbf{Log excerpt: R Agent} \\ \midrule

\texttt{[*] Security Review Session: 1} \\
\texttt{[*] Target Package: pyyaml-5.3} \\
\texttt{[*] Model: deepseek/deepseek-chat} \\
\texttt{[*] Temperature: 0.0, Max Tokens: 8000} \\
\texttt{[*] Grouping 2 query findings by MITRE data} \\
\texttt{[*] Group 'critical\_often\_medium': 2 queries,} \\
\texttt{~~~~6 results, 2 related CWEs} \\
\texttt{[+] Created 1 groups} \\
\texttt{[*] Phase 2: Review} \\
\texttt{[*] Reviewing group 'critical\_often\_medium' (2 queries)} \\
\texttt{[+] Starting review for group critical\_often\_medium.} \\
\texttt{[*] Iteration 1/10} \\[1ex]

\textbf{[DeepSeek]:} I'll start by examining the first critical finding in the \texttt{constructor.py} file at line 520. Let me extract the code snippet to understand the context. \\[1ex]

\texttt{[*] Tool call: extract\_code\_snippet} \\
\texttt{[*] Extracting lib/yaml/constructor.py:515-525} \\[1ex]

\textbf{[DeepSeek]:} I can see the critical issue now. The \texttt{find\_python\_name} function at line 540 contains \texttt{\_\_import\_\_(module\_name)} which is called when \texttt{unsafe=True}. This allows arbitrary module imports from YAML input. Let me check the \texttt{UnsafeConstructor} class to see how this is used. \\[1ex]

\texttt{[Calling tool: grep\_search]} \\
\texttt{[*] Iteration 10/10} \\[1ex]

\textbf{[DeepSeek]: [Tool Result]:} \\
\texttt{\{} \\
\texttt{~~"matches": [} \\
\texttt{~~~~\{"file": "lib3/yaml/constructor.py", "line": 704,} \\
\texttt{~~~~~"content": "class UnsafeConstructor(...):"\},} \\
\texttt{~~~~\{"file": "lib/yaml/constructor.py", "line": 716,} \\
\texttt{~~~~~"content": "class UnsafeConstructor(...):"\}],} \\
\texttt{~~"total": 2\}} \\[1ex]

\textbf{Current Status:} Iteration 10/10. (Remaining: 0) \\
\textbf{CRITICAL:} This is your FINAL iteration. You MUST submit all pending assessments. \\
\bottomrule
\end{tabularx}
\caption{R Agent execution log excerpt for pyyaml-5.3 vulnerability validation}
\label{tab:r-agent-log}
\end{table}

\begin{table}[htbp]
\footnotesize
\begin{tabularx}{\columnwidth}{@{}X@{}}
\toprule
\textbf{Log excerpt: S Agent} \\ \midrule

\texttt{[*] Sanity Check Session: 1} \\
\texttt{[*] Package: pyyaml-5.3} \\
\texttt{[*] Model: deepseek/deepseek-chat} \\
\texttt{[*] Temperature: 0.0, Max Tokens: 8000} \\
\texttt{[*] Validating finding 1/3: lib/yaml/constructor.py:520} \\
\texttt{[*] Extracting lib/yaml/constructor.py:510-530} \\[1ex]

\textbf{[+] LLM sanity check:} \\
\texttt{verdict=tp, confidence=95, recommendation=accept,} \\
\texttt{concerns=False, flags=ZERO\_DAY (critical): This appears} \\
\texttt{to be a legitimate, exploitable vulnerability in PyYAML's} \\
\texttt{default unsafe constructor} \\[1ex]

\texttt{[!] Finding 1 raised 1 flag(s): ZERO\_DAY} \\
\bottomrule
\end{tabularx}
\caption{S Agent execution log excerpt for pyyaml-5.3 finding validation}
\label{tab:s-agent-log}
\end{table}


\begin{table*}[!th]
\caption{Top100: CVE Detection Results and Comparison with Static Analysis Tools.}
\label{tab:top100-cve-detection}
\centering
\footnotesize
\setlength{\tabcolsep}{3pt}
\rowcolors{2}{gray!10}{white}
\begin{tabular}{@{}lllcccc@{}}
\toprule
CVE & Package & Configurations (Agreement) & Bandit & Semgrep & CodeQL  & Score\\
\midrule
\hiderowcolors
\textbf{Known CVEs (Rediscovered)} & & & & & \\
\showrowcolors
CVE-2020-13091 & pandas-2.3.3 & C$_{0,1}$, G$_{0,1}$, GPT$_{0,1}$, DC$_1$, DR$_{0,1}$ (9/10) & \checkmark & \checkmark & & 9.8\\
CVE-2019-6446 & numpy-2.3.4 & C$_{0,1}$, DC$_0$, G$_0$, GPT$_0$ (6/10) & \checkmark & \checkmark & & 9.8\\
CVE-2019-8341 & jinja2-3.1.6 & C$_{0,1}$, GPT$_{0,1}$, DR$_0$, G$_0$ (6/10) & & & &9.8 \\
CVE-2014-0012 & jinja2-3.1.6 & C$_{0,1}$, G$_{0,1}$, DC$_{0,1}$ (5/10) & \checkmark & \checkmark & &4.4\\
CVE-2014-1402 & jinja2-3.1.6 & C$_{0,1}$, G$_{0,1}$, DC$_{0,1}$ (5/10) & \checkmark & \checkmark & &4.4\\
CVE-2022-22817 & pillow-12.0.0 & C$_0$, G$_1$, GPT$_1$, DR$_1$ (4/10) & & & &9.8\\
CVE-2024-53899 & virtualenv-20.35.4 & DR$_{0,1}$, GPT$_{0,1}$ (4/10) & & \checkmark & & 7.8\\
CVE-2023-50447 & pillow-12.0.0 & C$_0$, GPT$_0$, DR$_1$ (3/10) & & & &8.1\\
CVE-2024-34062 & tqdm-4.67.1 & G$_0$, DC$_0$, GPT$_1$ (3/10) & \checkmark & \checkmark & &4.4\\
CVE-2014-3007 & pillow-12.0.0 & G$_{0,1}$, GPT$_0$ (3/10) & \checkmark & \checkmark & & 10\\
CVE-2017-18342 & pyyaml-6.0.3 & C$_0$, G$_1$ (2/10) & & & &9.8\\
CVE-2019-20477 & pyyaml-6.0.3 & C$_0$, G$_1$ (2/10) & & & & 9.8\\
CVE-2015-8557 & pygments-2.19.2 & GPT$_{0,1}$ (2/10) & \checkmark & & &9.8\\
CVE-2020-14343 & pyyaml-6.0.3 & C$_0$, G$_1$ (2/10) & & & &9.8\\
CVE-2020-1747 & pyyaml-6.0.3 & C$_0$, G$_1$ (2/10) & & & &9.8\\
CVE-2021-33503 & urllib3-2.5.0 & C$_0$, G$_0$ (2/10) & & & &7.5\\
CVE-2022-45061 & idna-3.11 & C$_{0,1}$ (2/10) & & & &7.5\\
CVE-2024-23334 & aiohttp-3.13.2 & C$_{0,1}$ (2/10) & & & & 7.5\\
CVE-2025-43859 & h11-0.16.0 & C$_{0,1}$ (2/10) & & & & 9.1\\
CVE-2014-1858 & numpy-2.3.4 & C$_0$ (1/10) & & & &5.5\\
CVE-2014-1859 & numpy-2.3.4 & C$_0$ (1/10) & & & &5.5\\
CVE-2016-1494 & rsa-4.9.1 & G$_0$ (1/10) & & & &5.3\\
CVE-2017-5992 & openpyxl-3.1.5 & C$_1$ (1/10) & & & & 8.2\\
CVE-2021-23437 & pillow-12.0.0 & C$_1$ (1/10) & & & &7.5\\
CVE-2021-27291 & pygments-2.19.2 & G$_0$ (1/10) & & & & 7.5\\
CVE-2022-29217 & pyjwt-2.10.1 & G$_0$ (1/10) & & & &7.5\\
CVE-2022-36087 & oauthlib-3.3.1 & DC$_1$ (1/10) & & & &6.5\\
CVE-2023-38325 & cryptography-46.0.3 & C$_0$ (1/10) & \checkmark & \checkmark & &7.5 \\
CVE-2024-5569 & zipp-3.23.0 & C$_0$ (1/10) & & & &6.2\\
\midrule
\hiderowcolors
\textbf{Newly Discovered CVEs} & & & & & \\
\showrowcolors
CVE-2025-68146 & filelock-3.20.0 & C$_{0,1}$, G$_0$, DC$_1$, DR$_1$, GPT$_0$ (6/10) & & & &6.3\\
CVE-2026-22701 & filelock-3.20.1 & C$_{0,1}$, G$_{0,1}$, DC$_1$, DR$_1$, GPT$_0$ (7/10) & & & &5.3\\
CVE-2026-22702 & virtualenv-20.36.0 & C$_1$ (1/10) & & & &4.5 \\
CVE-2026-23949 & setuptools-80.9.0 & C$_0$, G$_{0,1}$ (3/10) & \checkmark & & &8.6\\
CVE-2026-23490 & pyasn1-0.6.1 & C$_{0,1}$ (2/10) & & & &7.5\\
CVE-2026-34993 & aiohttp-3.13.2 & C$_1$, G$_{0,1}$, GPT$_{0,1}$, DR$_0$ (6/10) & \checkmark & \checkmark & & 7.3 \\
CVE-2026-28684 & python\_dotenv-1.2.1 & C$_{0,1}$ (2/10) & & & & 6.6 \\
CVE-2026-7246 & click-8.3.0 & C$_{0,1}$, GPT$_{0,1}$ (4/10) & & \checkmark & & 7.2 \\
\midrule
\hiderowcolors
\textbf{Acknowledgements (ACK)} & & & & & \\
\showrowcolors
DoS & jsonschema-4.25.1 & G$_0$, DR$_1$ (2/10) & & & & \\
TOCTOU & s3transfer-0.14.0 & G$_0$ (1/10) & & & & \\
Path Traversal & numpy-2.3.4 & C$_0$ (1/10) & & &&  \\
Code Injection & numpy-2.3.4 & C$_1$ (1/10) & \checkmark & \checkmark & & \\
\bottomrule
\end{tabular}
\end{table*}

\section{Statistics \& Metrics}
\label{sec:appendix-metric}
\input{package_stats}

\begin{figure*}[!ht]
    \centering
    \includegraphics[width=\linewidth]{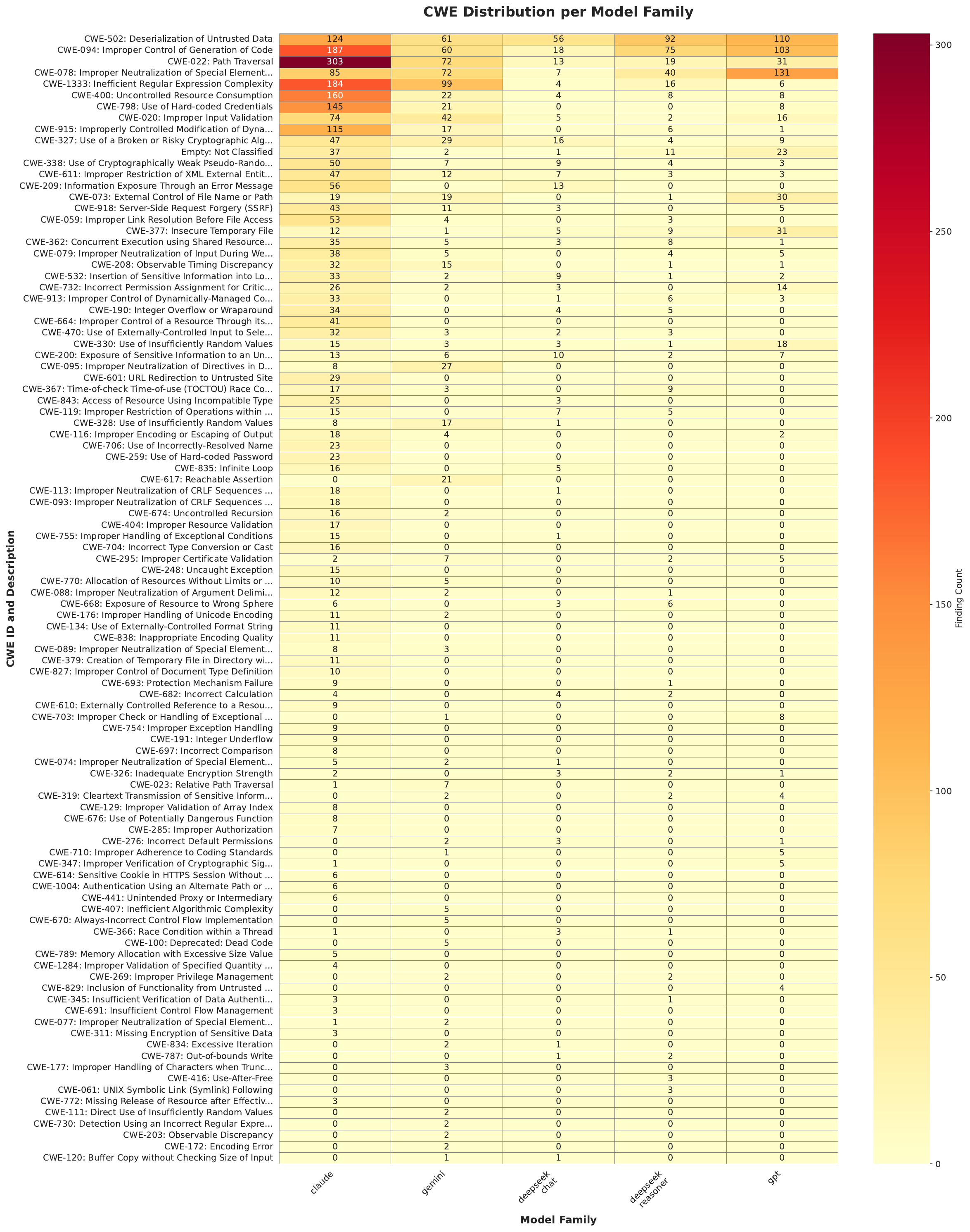}
    \caption{CWE Heatmap per model family.}
    \label{fig:cwe-heatmap}
\end{figure*}

\begin{figure*}[!ht]
    \centering
    \includegraphics[width=\linewidth]{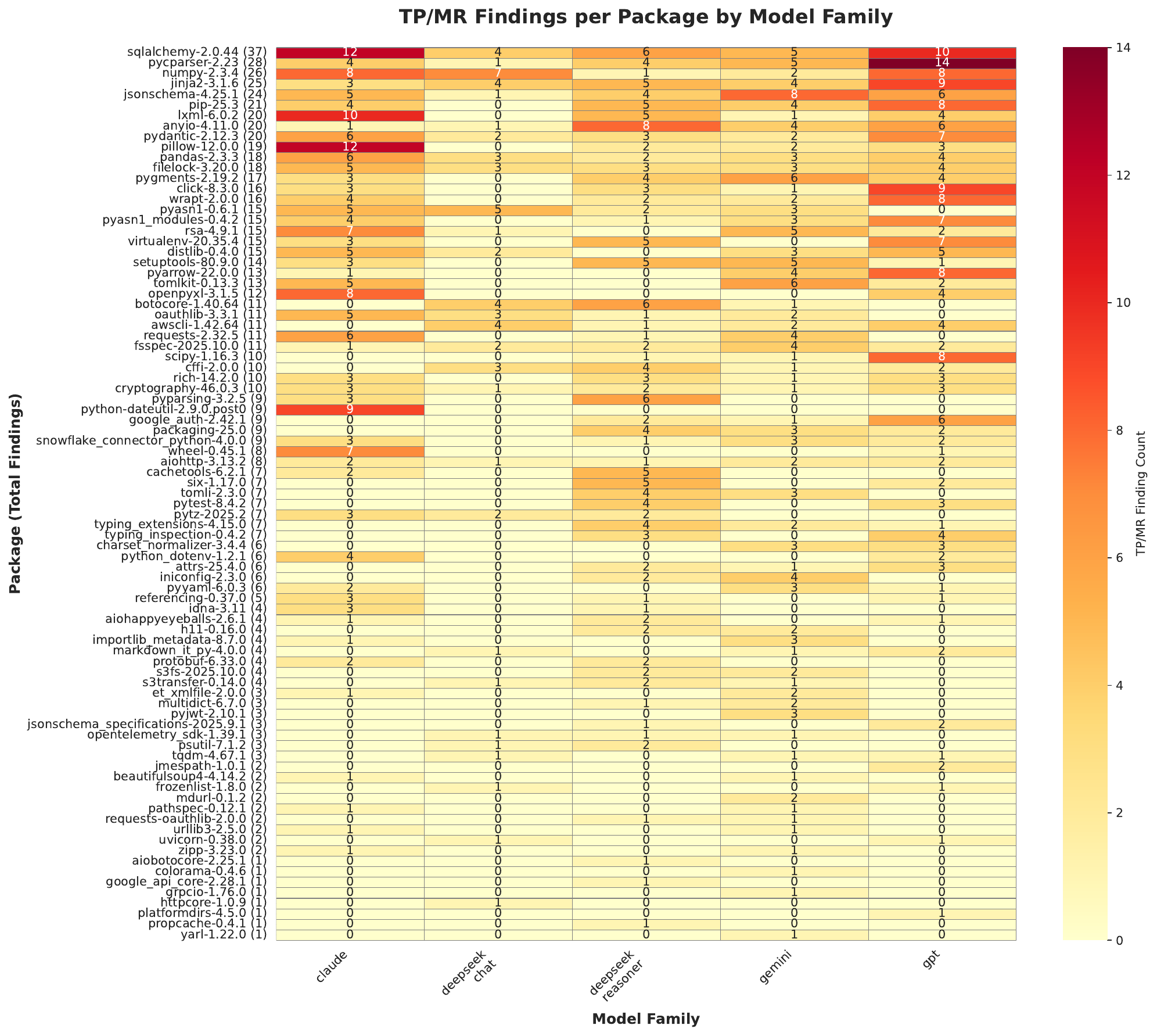}
    \caption{TP findings per package per model family.}
    \label{fig:findings-per-package}
\end{figure*}

\begin{figure*}[!ht]
    \centering
    \includegraphics[width=\linewidth]{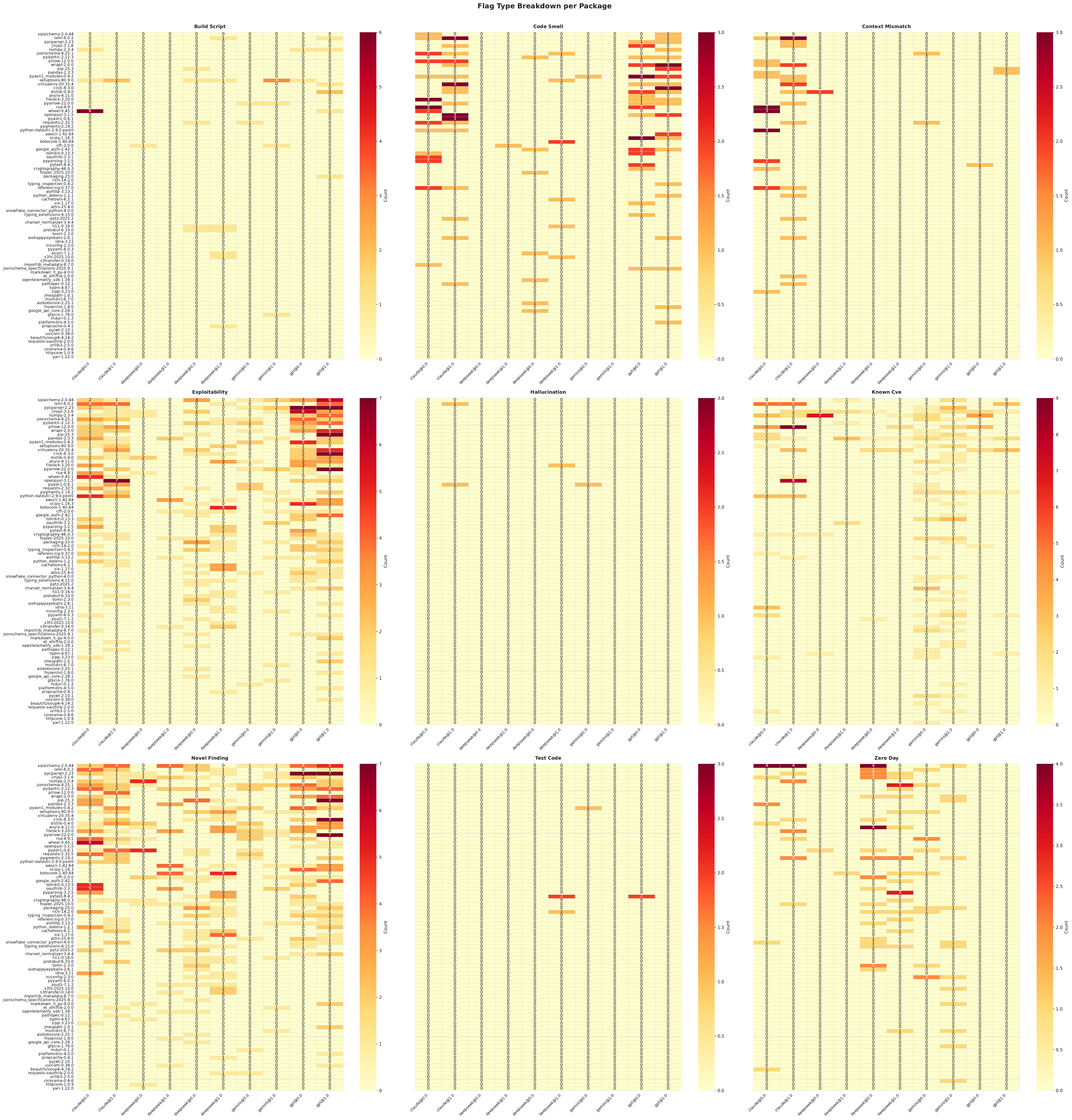}
    \caption{Package labels per model and temperature.}
    \label{fig:flags-per-package}
\end{figure*}

\begin{figure*}[!ht]
    \centering
    \includegraphics[width=\linewidth]{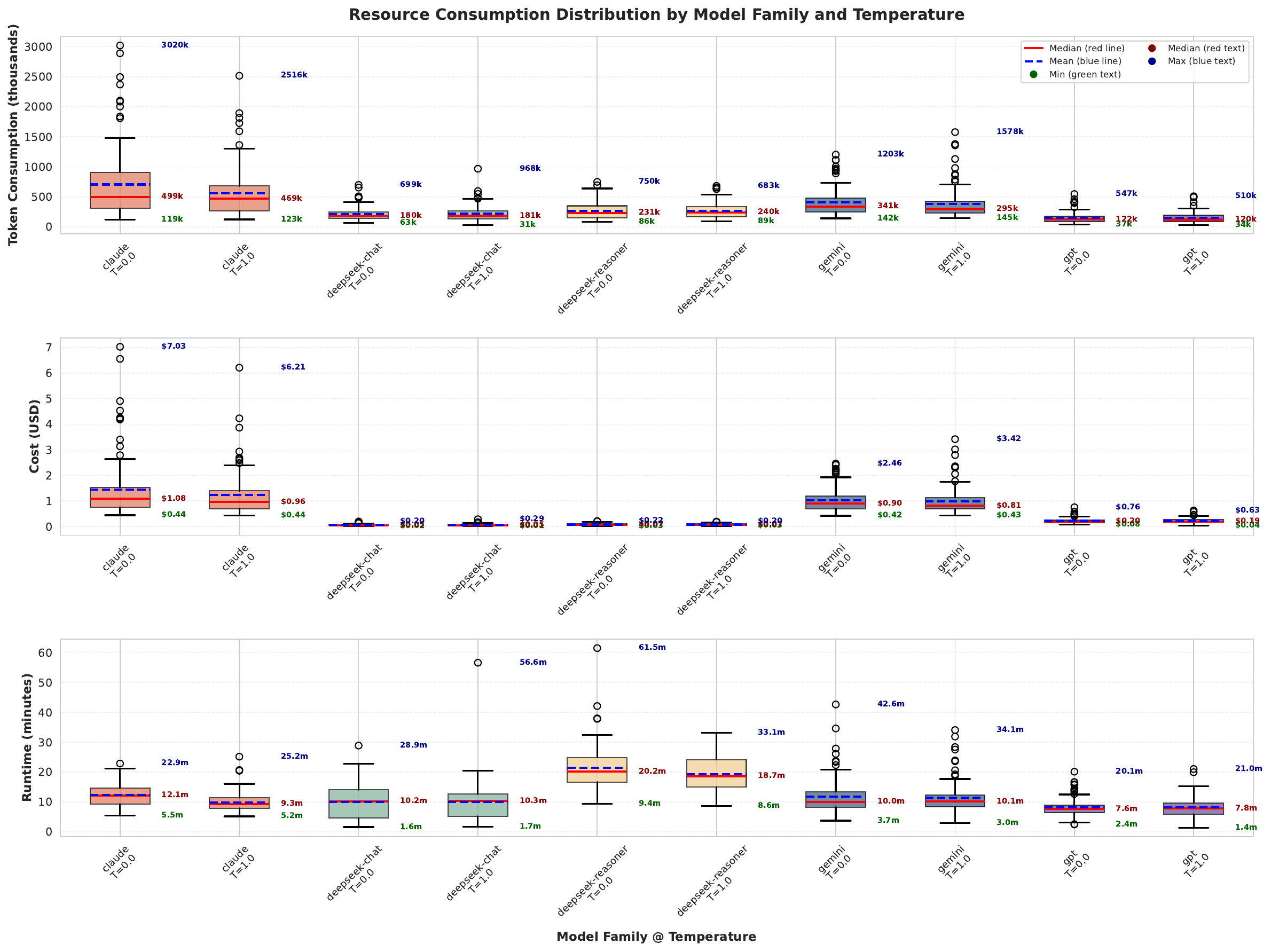}
    \caption{Resource metrics per model family.}
    \label{fig:resource-metrics}
\end{figure*}

\begin{figure*}[!ht]
    \centering
    \includegraphics[width=\linewidth]{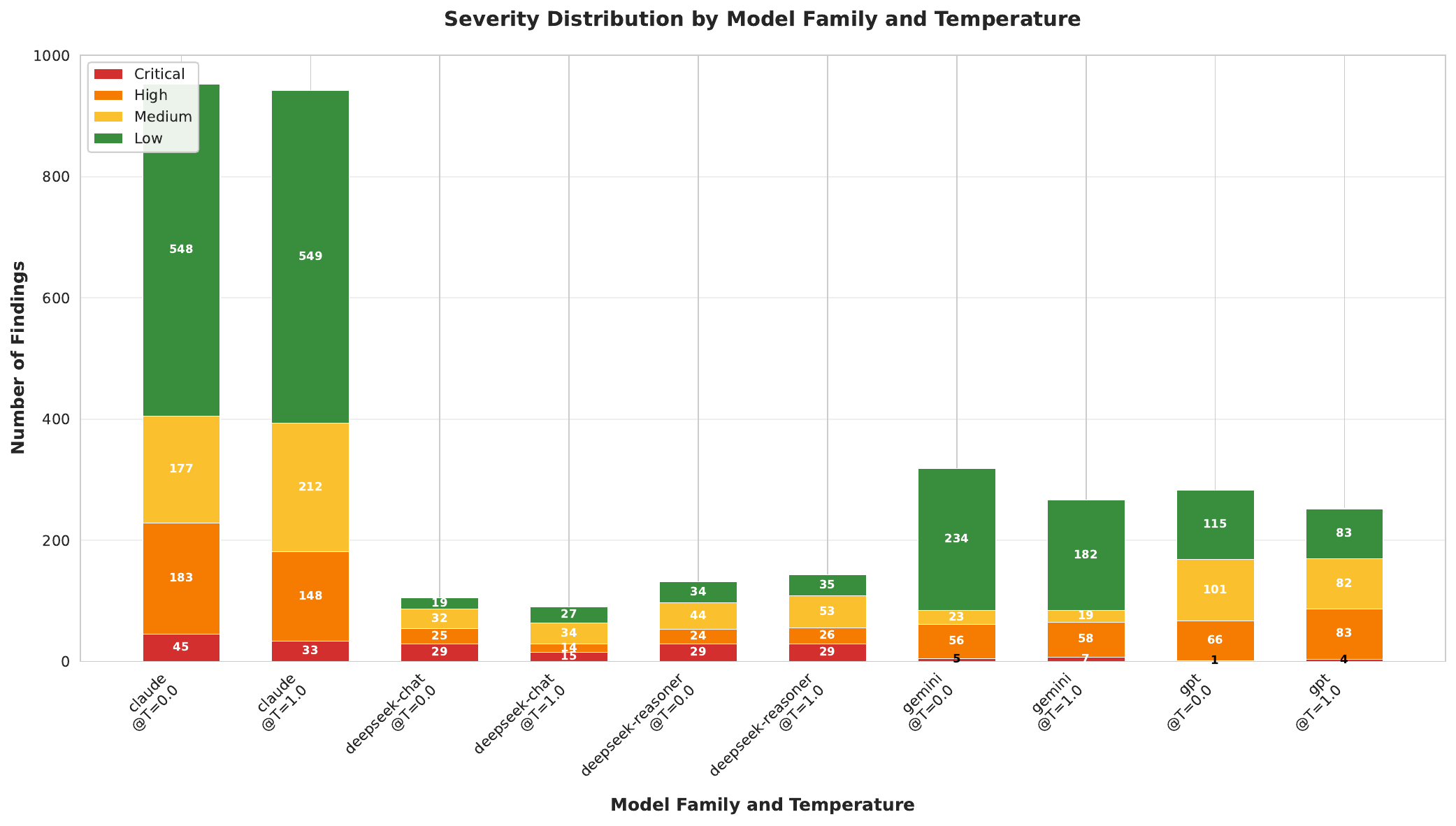}
    \caption{Severity distribution per model and temperature.}
    \label{fig:severity-distribution}
\end{figure*}

\begin{figure*}[!ht]
    \centering
    \includegraphics[width=\linewidth]{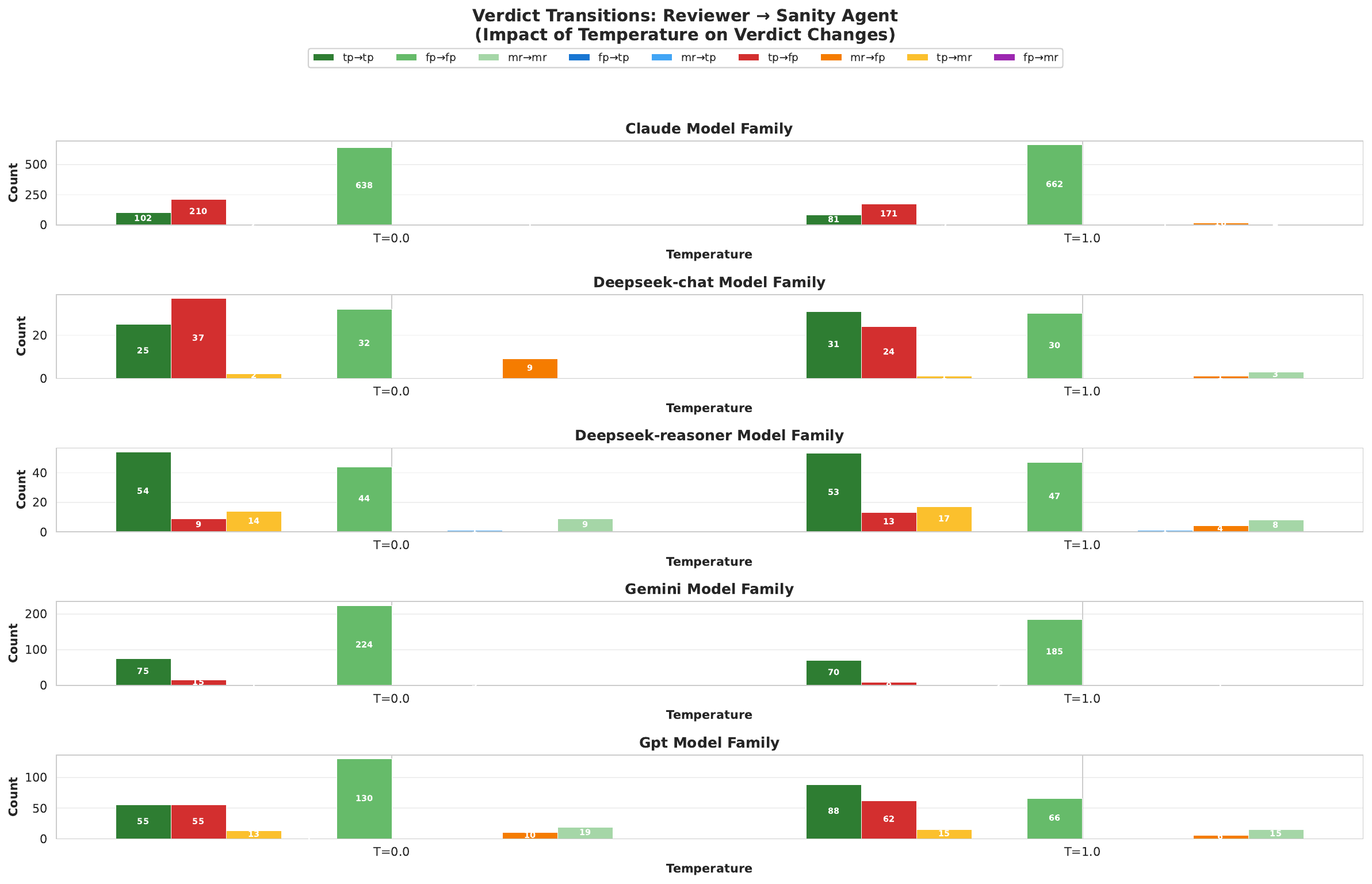}
    \caption{Verdict transitions per model family and temperature.}
    \label{fig:verdict-transitions}
\end{figure*}

\end{document}

%% file: architecture.tex
\begin{tikzpicture}[
  font=\sffamily,
  >=Stealth,
  node distance=10mm and 25mm,
  agent/.style={
    rounded corners=3pt,
    fill=geoDBlue,
    draw=geoDBlue,
    text=white,
    minimum width=28mm,
    minimum height=10mm,
    font=\bfseries,
    align=center,
    line width=0.9pt,
    drop shadow={opacity=0.12,shadow xshift=0.5pt,shadow yshift=-0.5pt}
  },
  doc/.style={
    shape=tape,
    draw=neutralGray,
    fill=white,
    minimum width=26mm,
    minimum height=17mm,
    align=center,
    inner sep=3mm,
    line width=0.8pt,
    drop shadow={opacity=0.12,shadow xshift=0.5pt,shadow yshift=-0.5pt},
    tape bend top=none
  },
  toolpanel/.style={
    draw=geoMBlue,
    rounded corners=3pt,
    fill=geoLBlue,
    inner sep=3mm,
    minimum width=46mm,
    align=left,
    line width=0.7pt,
    drop shadow={opacity=0.1,shadow xshift=0.5pt,shadow yshift=-0.5pt}
  },
  paneltitle/.style={
    font=\small\bfseries,
    inner sep=1.5mm,
    fill=geoMBlue,
    draw=geoMBlue,
    rounded corners=2pt,
    line width=0.6pt
  },
  statpanel/.style={
    draw=neutralGray!90,
    rounded corners=3pt,
    fill=lightGray!20,
    inner sep=2.5mm,
    minimum width=42mm,
    align=left,
    line width=0.7pt,
    drop shadow={opacity=0.1,shadow xshift=0.5pt,shadow yshift=-0.5pt}
  },
  stattitle/.style={
    font=\small\bfseries,
    inner sep=1.5mm,
    fill=neutralGray!30,
    draw=neutralGray!90,
    rounded corners=2pt,
    line width=0.6pt
  },
  lbl/.style={font=\small, text=darkGray},
  flow/.style={->, draw=primaryBlue, line width=1.2pt},
  feed/.style={->, draw=neutralGray!80, line width=0.9pt, dashed},
  loop/.style={
    ->,
    draw=geoDBlue,
    line width=1.2pt,
    rounded corners=3pt,
    dotted
  }
]

\node[doc] (input) {Preprompt\\High-Level\\Requirements};
\node[agent, right=of input] (Q) {\textsc{Q Agent}};
\node[agent, right=of Q] (R) {\textsc{R Agent}};
\node[agent, right=of R] (S) {\textsc{S Agent}};
\node[doc, right=of S] (report) {Final\\Report};

\node[doc, above=of Q, yshift=5mm] (know) {\textbf{Knowledge}\\Base};

\draw[flow] (input) -- node[lbl, above]{optional} (Q);
\draw[flow] (Q) -- node[lbl, above]{findings} (R);
\draw[flow] (R) -- node[lbl, above]{verified} (S);
\draw[flow] (S) -- (report);
\draw[feed] (know) -- node[lbl, right, align=center]{feed} (Q);

\node[statpanel, above=of R, yshift=3mm] (statR) {\small\texttt{group\_findings\_by\_mitre}};
\node[stattitle, anchor=south west] at (statR.north west) {Static Tools};
\draw[feed] (statR) -- node[lbl, right, align=center]{clustering} (R);

\node[statpanel, above=of S, yshift=3mm] (statS) {\small\texttt{search\_cve}};
\node[stattitle, anchor=south west] at (statS.north west) {Static Tools};
\draw[feed] (statS) -- node[lbl, right, align=center]{online\\search} (S);

\node[toolpanel, below=of Q, yshift=-3mm] (toolsQ) {
  \small\texttt{get\_package\_profile}\\[1pt]
  \small\texttt{execute\_query}\\[1pt]
  \small\texttt{fixup\_query}\\[1pt]
  \small\texttt{create\_report}
};
\node[paneltitle, anchor=south west] at (toolsQ.north west) {LLM Tools};

\node[toolpanel, below=of R, yshift=-3mm] (toolsR) {
  \small\texttt{extract\_code\_snippet}\\[1pt]
  \small\texttt{grep\_search}\\[1pt]
  \small\texttt{trace\_flow}\\[1pt]
  \small\texttt{submit\_finding}
};
\node[paneltitle, anchor=south west] at (toolsR.north west) {LLM Tools};

\node[toolpanel, below=of S, yshift=-3mm] (toolsS) {
  \small\texttt{verify\_finding}\\[1pt]
  \small\texttt{flag\_finding}
};
\node[paneltitle, anchor=south west] at (toolsS.north west) {LLM Tools};

\draw[feed] (Q) -- (toolsQ.north);
\draw[feed] (R) -- (toolsR.north);
\draw[feed] (S) -- (toolsS.north);

\begin{scope}[on background layer]
\draw[loop]
  (toolsQ.south) -- ++(0,-5mm)
  coordinate (q1) -- ++(-20mm,0)
  coordinate (q2) |- (Q.south west);
\node[font=\scriptsize\itshape, text=geoDBlue] at ($(q1)!0.5!(q2)+(0,-2mm)$) {agentic loop};

\draw[loop]
  (toolsR.south) -- ++(0,-5mm)
  coordinate (r1) -- ++(-20mm,0)
  coordinate (r2) |- (R.south west);
\node[font=\scriptsize\itshape, text=geoDBlue] at ($(r1)!0.5!(r2)+(0,-2mm)$) {agentic loop};

\draw[loop]
  (toolsS.south) -- ++(0,-5mm)
  coordinate (s1) -- ++(-20mm,0)
  coordinate (s2) |- (S.south west);
\node[font=\scriptsize\itshape, text=geoDBlue] at ($(s1)!0.5!(s2)+(0,-2mm)$) {iterative loop};
\end{scope}

\end{tikzpicture}

%% file: package_stats.tex
\begin{table*}
\caption{Top 100 PyPI Packages: Metadata and Popularity Metrics}
\label{tab:top100-packages}
\tiny
\begin{tabular}{@{}lrrrrrrr@{}}
\toprule
\textbf{Package-Version} & \textbf{Downloads} & \textbf{Commits} & \textbf{Used By} & \textbf{Contributors} & \textbf{Stars} & \textbf{Files} & \textbf{Size} \\
\midrule
aiobotocore-2.25.1 & 882.7M & 1.1k & 14.9k & 76 & 1.4k & 68 & 481.9KB \\
aiohappyeyeballs-2.6.1 & 218.0M & 208 & 116.1k & 13 & 37 & 19 & 117.4KB \\
aiohttp-3.13.2 & 280.8M & 12.9k & 752.3k & 804 & 16.2k & 370 & 15.5MB \\
aiosignal-1.4.0 & 245.7M & 762 & N/A & 12 & 161 & 36 & 78.8KB \\
annotated\_types-0.7.0 & 371.1M & 83 & 400.7k & 20 & 588 & 18 & 62.6KB \\
anyio-4.11.0 & 395.1M & 1.2k & 610.7k & 69 & 2.4k & 114 & 1,012.6KB \\
attrs-25.4.0 & 442.6M & 1.7k & 1.4M & 182 & 5.7k & 129 & 1.9MB \\
awscli-1.42.64 & 346.5M & 13.6k & 5 & 481 & 16.7k & 6k & 9.5MB \\
beautifulsoup4-4.14.2 & 207.8M & N/A & N/A & N/A & N/A & 87 & 2.3MB \\
boto3-1.40.64 & 1,438.5M & 7.4k & 577.0k & 168 & 9.7k & 71 & 979.3KB \\
botocore-1.40.64 & 1,010.4M & 12.8k & 423.9k & 218 & 1.6k & 3k & 115.4MB \\
cachetools-6.2.1 & 357.6M & 399 & 680.0k & 19 & 2.7k & 35 & 220.8KB \\
certifi-2025.10.5 & 955.3M & 327 & 1 & 40 & 952 & 17 & 297.8KB \\
cffi-2.0.0 & 541.8M & 3.6k & N/A & 96 & 234 & 200 & 2.1MB \\
charset\_normalizer-3.4.4 & 890.6M & 726 & 1.5M & 24 & 732 & 80 & 386.9KB \\
click-8.3.0 & 483.1M & 3.0k & 2.3M & 421 & 17.1k & 103 & 1.4MB \\
colorama-0.4.6 & 289.5M & 391 & 1.4M & 52 & 3.8k & 31 & 113.0KB \\
cryptography-46.0.3 & 581.2M & 14.6k & 920.0k & 360 & 7.4k & 441 & 4.0MB \\
distlib-0.4.0 & 208.0M & 1.3k & N/A & 41 & 60 & 194 & 1.9MB \\
et\_xmlfile-2.0.0 & 211.3M & 119 & N/A & N/A & N/A & 15 & 63.1KB \\
fastapi-0.120.4 & 205.7M & 6.6k & 840.4k & 872 & 94.4k & 1k & 3.3MB \\
filelock-3.20.0 & 351.6M & 417 & 695.5k & 55 & 927 & 29 & 139.7KB \\
frozenlist-1.8.0 & 254.2M & 932 & 362.9k & 20 & 118 & 45 & 155.5KB \\
fsspec-2025.10.0 & 560.0M & 2.5k & 282.9k & 305 & 1.3k & 92 & 949.1KB \\
google\_api\_core-2.28.1 & 293.6M & 565 & N/A & 84 & 145 & 130 & 1.0MB \\
google\_auth-2.42.1 & 335.8M & 1.0k & N/A & 152 & 868 & 180 & 1.7MB \\
googleapis\_common\_protos-1.71.0 & 263.1M & 40.9k & N/A & 544 & 5.2k & 205 & 740.1KB \\
greenlet-3.2.4 & 250.4M & 1.2k & 586.6k & 84 & 1.8k & 133 & 654.1KB \\
grpcio\_status-1.76.0 & 748.3M & N/A & N/A & N/A & N/A & 18 & 44.7KB \\
grpcio-1.76.0 & 257.2M & 57.9k & 6.6k & 1.1k & 44.3k & 5k & 74.0MB \\
h11-0.16.0 & 396.7M & 482 & 646.4k & 34 & 546 & 69 & 422.5KB \\
httpcore-1.0.9 & 295.2M & 567 & 395.0k & 59 & 529 & 61 & 490.6KB \\
httpx-0.28.1 & 300.4M & 1.5k & 551.0k & 246 & 14.9k & 70 & 829.3KB \\
idna-3.11 & 909.3M & 390 & 2.4M & 28 & 274 & 26 & 1.9MB \\
importlib\_metadata-8.7.0 & 365.2M & 1.8k & N/A & 67 & 139 & 76 & 260.0KB \\
iniconfig-2.3.0 & 372.2M & 190 & 595.2k & 16 & 58 & 23 & 62.3KB \\
jinja2-3.1.6 & 370.6M & 2.9k & 19 & 335 & 11.4k & 114 & 1.5MB \\
jmespath-1.0.1 & 405.9M & 578 & N/A & 32 & 2.4k & 22 & 105.4KB \\
jsonschema\_specifications-2025.9.1 & 231.7M & 295 & 332.1k & 6 & 12 & 45 & 117.1KB \\
jsonschema-4.25.1 & 270.8M & 3.0k & 808.5k & 120 & 4.9k & 643 & 3.2MB \\
lxml-6.0.2 & 210.3M & 6.8k & 583.8k & 173 & 3.0k & 466 & 30.5MB \\
markdown\_it\_py-4.0.0 & 249.6M & 342 & 349.4k & 31 & 1.2k & 123 & 383.3KB \\
markupsafe-3.0.3 & 408.2M & 844 & 2.2M & 49 & 685 & 37 & 256.1KB \\
mdurl-0.1.2 & 241.8M & 37 & 342.7k & 4 & 14 & 17 & 43.8KB \\
multidict-6.7.0 & 270.8M & 1.6k & N/A & 61 & 480 & 113 & 598.7KB \\
numpy-2.3.4 & 658.8M & 40.6k & 3.5M & 2.0k & 31.3k & 8k & 63.4MB \\
oauthlib-3.3.1 & 229.4M & 1.8k & 641.5k & 202 & 3.0k & 222 & 1.3MB \\
openpyxl-3.1.5 & 210.3M & 9.1k & N/A & N/A & 11 & 377 & 2.0MB \\
opentelemetry\_sdk-1.39.1 & 246.0M & 2.0k & N/A & 330 & 2.3k & 131 & 1.1MB \\
packaging-25.0 & 860.0M & 893 & 1.8M & 131 & 709 & 79 & 2.4MB \\
pandas-2.3.3 & 497.0M & 37.1k & 2.7M & 4.1k & 47.7k & 2k & 21.7MB \\
pathspec-0.12.1 & 241.4M & 362 & N/A & 26 & 209 & 33 & 212.1KB \\
pillow-12.0.0 & 239.6M & 20.3k & 2.4M & 493 & 13.3k & 2k & 69.2MB \\
pip-25.3 & 429.5M & 15.7k & 242.7k & 869 & 10.1k & 833 & 9.2MB \\
platformdirs-4.5.0 & 405.7M & 548 & 777.1k & 80 & 885 & 21 & 124.9KB \\
pluggy-1.6.0 & 467.1M & 924 & 765.6k & 54 & 1.6k & 70 & 233.4KB \\
propcache-0.4.1 & 244.3M & 173 & N/A & 12 & 33 & 58 & 158.6KB \\
protobuf-6.33.0 & 480.2M & 22.3k & 957.0k & 1.4k & 70.4k & 273 & 2.7MB \\
psutil-7.1.2 & 239.2M & 6.5k & 753.6k & 228 & 11.0k & 200 & 2.1MB \\
pyarrow-22.0.0 & 263.2M & 18.5k & 4 & 1.5k & 16.4k & 304 & 5.6MB \\
pyasn1\_modules-0.4.2 & 306.1M & 351 & N/A & 16 & 8 & 286 & 1.4MB \\
pyasn1-0.6.1 & 367.1M & 1.0k & 18.1k & 13 & 47 & 178 & 1.2MB \\
pycparser-2.23 & 477.1M & 651 & 1.1M & 91 & 3.5k & 206 & 1.2MB \\
pydantic\_core-2.41.4 & 441.8M & 1.5k & N/A & 111 & 1.7k & 254 & 2.8MB \\
pydantic-2.12.3 & 549.0M & 5.4k & 903.2k & 724 & 26.6k & 288 & 4.0MB \\
pygments-2.19.2 & 463.6M & 7.1k & 1.1M & 892 & 2.1k & 3k & 43.1MB \\
pyjwt-2.10.1 & 311.0M & 933 & 789.6k & 163 & 5.6k & 86 & 337.6KB \\
pyparsing-3.2.5 & 228.0M & 1.7k & 1.3M & 75 & 2.4k & 205 & 4.1MB \\
pytest-8.4.2 & 396.0M & 17.1k & 1.6M & 1.1k & 13.5k & 617 & 5.9MB \\
python\_dotenv-1.2.1 & 273.1M & 415 & 1.6M & 109 & 8.6k & 48 & 188.1KB \\
python-dateutil-2.9.0.post0 & 717.6M & 1.6k & N/A & 133 & 2.6k & 98 & 1.0MB \\
pytz-2025.2 & 377.3M & 679 & 2.5M & 21 & 384 & 624 & 1.0MB \\
pyyaml-6.0.3 & 575.5M & 326 & 1.7M & 40 & 2.8k & 640 & 600.3KB \\
referencing-0.37.0 & 236.1M & 799 & 332.3k & 6 & 52 & 353 & 368.7KB \\
requests-2.32.5 & 959.1M & 6.4k & 4.0M & 776 & 53.7k & 84 & 469.1KB \\
requests-oauthlib-2.0.0 & 264.5M & 666 & 595.6k & 119 & 1.8k & 63 & 225.8KB \\
rich-14.2.0 & 264.9M & 4.4k & 473.8k & 285 & 55.2k & 83 & 950.2KB \\
rpds\_py-0.28.0 & 242.0M & 543 & N/A & 20 & 59 & 31 & 117.2KB \\
rsa-4.9.1 & 401.3M & 604 & N/A & 43 & 492 & 20 & 105.0KB \\
s3fs-2025.10.0 & 572.3M & 1.1k & 25.5k & 151 & 1.0k & 33 & 319.1KB \\
s3transfer-0.14.0 & 654.3M & 720 & 372.9k & 44 & 232 & 62 & 793.3KB \\
scipy-1.16.3 & 262.2M & 36.8k & 1.4M & 1.9k & 14.4k & 5k & 107.2MB \\
setuptools-80.9.0 & 880.4M & 16.5k & 805.2k & 635 & 2.8k & 582 & 4.9MB \\
six-1.17.0 & 674.3M & 548 & 2.7M & 67 & 1.0k & 16 & 131.2KB \\
sniffio-1.3.1 & 184.3M & 83 & 676.6k & 16 & 147 & 38 & 54.2KB \\
snowflake\_connector\_python-4.0.0 & 144.7M & 1.8k & N/A & 145 & 705 & 336 & 9.9MB \\
soupsieve-2.8 & 203.0M & 371 & 736.3k & 13 & 263 & 134 & 515.2KB \\
sqlalchemy-2.0.44 & 250.3M & 17.9k & 1.2M & 809 & 11.4k & 1k & 60.6MB \\
tomli-2.3.0 & 256.7M & 325 & 614.3k & 13 & 550 & 9 & 54.6KB \\
tomlkit-0.13.3 & 202.3M & 454 & 156.0k & 45 & 807 & 809 & 2.1MB \\
tqdm-4.67.1 & 261.3M & 2.0k & 1.1M & 118 & 30.9k & 98 & 682.9KB \\
typing\_extensions-4.15.0 & 990.3M & 504 & N/A & 78 & 547 & 10 & 710.2KB \\
typing\_inspection-0.4.2 & 311.2M & 48 & 73.2k & 3 & 57 & 32 & 257.9KB \\
urllib3-2.5.0 & 1,106.1M & 4.3k & 2.4M & 385 & 4.0k & 154 & 1.9MB \\
uvicorn-0.38.0 & 217.8M & 1.4k & 828.6k & 209 & 10.3k & 79 & 439.1KB \\
virtualenv-20.35.4 & 273.5M & 1.0k & 492.5k & 149 & 5.0k & 188 & 6.4MB \\
wheel-0.45.1 & 322.2M & 994 & 370.7k & 86 & 547 & 102 & 476.4KB \\
wrapt-2.0.0 & 256.8M & 1.1k & 610.1k & 36 & 2.3k & 134 & 678.2KB \\
yarl-1.22.0 & 271.5M & 1.9k & N/A & 100 & 1.5k & 72 & 660.1KB \\
zipp-3.23.0 & 326.9M & 895 & 880.8k & 32 & 67 & 45 & 74.9KB \\
\bottomrule
\end{tabular}
\end{table*}